\documentclass[journal,comsoc]{IEEEtran}

\usepackage{array,graphicx,epsfig,amssymb,amsfonts}
\usepackage{bbm,amsmath,epstopdf,algorithm,algorithmic,subfigure, multirow}
\usepackage{comment}
\usepackage{color}
\usepackage{cases}
\usepackage{slashbox}
\usepackage{blkarray}

\usepackage{pifont}
\newcommand{\cmark}{\ding{51}}%
\newcommand{\xmark}{\ding{55}}%

\newcommand{\be}{\begin{equation}}
\newcommand{\ee}{\end{equation}}
\def\bee#1\eee{\begin{align}#1\end{align}}
\newcommand{\bse}{\begin{subequations}}
\newcommand{\ese}{\end{subequations}}
\newcommand{\nnb}{\nonumber}

\newtheorem{theorem}{\textbf{Theorem}}

\newtheorem{lemma}{\textbf{Lemma}}
\newtheorem{corollary}{\textbf{Corollary}}

\newtheorem{definition}{\textbf{Definition}}

\def\prob{\mathsf{Prob}}

\newcommand{\red}[1]{{\color{red}#1}}

\newcommand{\bm}[1]{{\boldsymbol#1}}

\ifodd 0
\newcommand{\rev}[1]{{{\color{blue} #1}}} 
\else
\newcommand{\rev}[1]{#1}
\fi

\def \ISTR {}

\begin{document}

\title{Delay-Constrained Input-Queued Switch}

\author{Lei~Deng,~\IEEEmembership{Member,~IEEE,}
        Wing~Shing~Wong,~\IEEEmembership{Fellow,~IEEE,}
        Po-Ning~Chen,~\IEEEmembership{Senior Member,~IEEE,}
        Yunghsiang~S.~Han,~\IEEEmembership{Fellow,~IEEE,}
        and Hanxu Hou,~\IEEEmembership{Member,~IEEE}
\thanks{This work was partially supported by Schneider Electric,  Lenovo Group (China) Limited and the Hong Kong Innovation and Technology Fund (ITS/066/17FP)
under the HKUST-MIT Research Alliance Consortium,
by the Research Grants Council of the Hong Kong Special Administrative Region under Project GRF 14200217,
and by the NSFC of China (No.~61671007 and No. 61701115). A two-page preliminary version of this paper was published as a poster paper
in ACM MobiHoc 2018 \cite{delaydeng2018}.
}
\thanks{L.~Deng, Y.~S.~Han (corresponding author) and H.~Hou are with
School of Electrical Engineering \& Intelligentization,
Dongguan University of Technology (email: denglei@dgut.edu.cn, yunghsiangh@gmail.com, houhx@dgut.edu.cn).}
\thanks{W.~S.~Wong is with Department of Information Engineering, The
Chinese University of Hong Kong (email: wswong@ie.cuhk.edu.hk).}
\thanks{P.-N.~Chen is with Department of Electrical and Computer
Engineering, National Chiao Tung University (email: poning@faculty.nctu.edu.tw).}}

\maketitle

\begin{abstract}
In this paper, we study the delay-constrained input-queued switch where each packet has a deadline and it will expire
if it is not delivered before its deadline. Such new scenario is motivated by the proliferation of real-time applications
in multimedia communication systems,  tactile Internet, networked controlled systems, and cyber-physical systems.
The delay-constrained input-queued switch is completely different from the well-understood delay-unconstrained one
and thus poses new challenges. We focus on three fundamental problems centering around the performance metric of \emph{timely throughput}:
(i) how to characterize the capacity region? (ii) how to design a feasibility/throughput-optimal scheduling policy? and (iii) how
to design a network-utility-maximization scheduling policy? We use three different approaches to solve these three fundamental problems.
The first approach is based on  Markov Decision Process (MDP) theory, which can solve all three problems. However,
it suffers from the curse of dimensionality. The second approach breaks the curse of dimensionality by exploiting
the combinatorial features of the problem. It gives a new capacity region characterization with only a polynomial number of linear constraints.
The third approach is based on the framework of Lyapunov optimization, where we design a polynomial-time
maximum-weight $T$-disjoint-matching scheduling policy which is proved to be feasibility/throughput-optimal.
Our three approaches apply to the frame-synchronized traffic pattern but
our MDP-based approach can be extended to more general traffic patterns.
\end{abstract}
\section{Introduction}
Switches, which interconnect multiple devices, are the core  of communication networks.
There are mainly three types of switch designs: output-queued switch,  direct input-queued switch,
and  input-queued switch using virtual output queueing.
Among them, the input-queued switch using virtual output queueing is most widely used because
it addresses the $N$-speedup problem of the output-queued switch \cite{chuang1999matching,kang2013design} and the Head-Of-Line (HOL) blocking problem of
the direct input-queued switch \cite{karol1987input}.
In this work, we study the input-queued switch using virtual output queueing, which we simply call \emph{input-queued switch} for the sake of convenience.

Most existing works on input-queued switches consider \emph{delay-unconstrained} traffic where packets can be kept in the virtual output queues forever.
Throughput and average delay are two major performance metrics for delay-unconstrained input-queued switches.
The authors in \cite{mckeown1999achieving} characterized the capacity region for independent, identically distributed (i.i.d.) arrivals and
further proved that the maximum-weight-matching scheduling policy is \emph{throughput-optimal} in the sense that it can support any feasible throughput requirements in the capacity region.
The authors in \cite{dai2000throughput} extended these results to arbitrary delay-unconstrained arrivals by using fluid model techniques.
To study the average delay performance, the authors in \cite{neely2007logarithmic} proposed another throughput-optimal scheduling policy and
showed that it attains $O(\log N)$ average delay for $N \times N$ input-queued switches.

However,  with the proliferation of real-time applications, the communication networks nowadays need to support more and more \emph{delay-constrained} traffic. Typical examples include multimedia communication systems such
as real-time streaming and video conferencing \cite{deng2017timely}, tactile Internet \cite{fettweis2014tactile,simsek20165g},
networked controlled systems (NCSs) such as remote control of unmanned aerial vehicles (UAVs) \cite{baillieul2007control,zeyu2016autonomous},
and cyber-physical systems (CPSs) such as medical tele-operations, X-by-wire vehilces/avionics, factory automation, and robotic collaboration \cite{kang2013design}.
In such applications, each packet has a hard deadline: if it is not delivered before its deadline, its validity will expire and it will be removed from the system.
In addition, throughput, which is termed \emph{timely throughput} in the delay-constrained scenario \cite{hou2009qos,hou2010utility,kang2016performance,deng2017timely}, is also important to such applications.
Taking NCSs as an example, the control system can be stabilized
if the control messages arrive before the predetermined deadlines and the dropout rate is below a threshold
(which equivalently means that the timely throughput is above a threshold) \cite{tan2017necessary, tan2017delay}.
Taking tactile Internet as another example, the timely throughput is a measure of reliability \cite{simsek20165g}.

\begin{table*}[t]
\centering
\caption{Our three approaches to solving the three fundamental problems for delay-constrained input-queued switches.}
\label{tab:result-summary}
\begin{tabular}{|c|c|c|c|c|c|}
\hline
Approach       & Capacity Region & \begin{tabular}[c]{@{}c@{}}Feasibility/Throughput-\\ Optimal Scheduling Policy\end{tabular} & \begin{tabular}[c]{@{}c@{}}Network-Utility-Maxi.\\ Scheduling Policy\end{tabular} & Complexity & \begin{tabular}[c]{@{}c@{}}Extend to General\\ Traffic Pattern\end{tabular}  \\ \hline
MDP-based (Sec.~\ref{sec:MDP})   & \cmark          & \cmark                                                                                     &      \cmark                                                                              & Exponential       & \cmark \\ \hline
Combinatorial (Sec.~\ref{sec:capacity-region}) &  \cmark         &  \xmark                                                                                    &       \xmark                                                                             & Polynomial           & \xmark \\ \hline
Lyapunov-based (Sec.~\ref{sec:Lyapunov}) &  \xmark         &  \cmark                                                                                    &       \xmark                                                                             & Polynomial        & \xmark \\ \hline
\end{tabular}
\vspace{-0.2cm}
\end{table*}

Since switches are the core of communication networks, how to support delay-constrained traffic in switches becomes critical.
Note that switches can serve delay-constrained traffi
c such as tactile applications from both wireless ends and wireline ends.
There are some existing works that investigate how to design real-time input-queued switch, e.g., \cite{chang2006providing,wang2008switch,kang2013design}.
In \cite{chang2006providing}, the authors proposed two scheduling policies under which the delivery delay
of packets is upper bounded by a finite value.
In \cite{wang2008switch,kang2013design},
the design goal is to deliver \emph{all} packets and minimize the maximum delivery delay among all packets.
Thus, existing works do not directly guarantee the delivery of delay-constrained traffics where hard deadlines are predetermined by the applications; and they do not allow any packet loss.
Instead, in this work, we consider how to deliver delay-constrained traffic and focus on the performance metric of timely throughput. More specifically,
we study the following three fundamental problems for delay-constrained input-queued switches:
\begin{itemize}
\item First, we aim to characterize the capacity region
in terms of timely throughput of all input-output pairs. The capacity region serves as the foundation to evaluate the performance of
any scheduling policy.
\item Second, we aim to design a throughput-optimal (which is termed \emph{feasibility-optimal} in the delay-constrained scenario \cite{hou2009qos,hou2010utility, kang2016performance,deng2017timely}) scheduling policy,
which can support any feasible timely throughput requirements in the capacity region. This problem is important for inelastic applications which have stringent minimum timely throughput requirements.
\item Third, we aim to design a scheduling policy to maximize the network utility with respect to the achieved timely throughput. This problem
is important for elastic applications which do not have stringent minimum timely throughput requirements but aim to obtain large utility.
Here an elastic application has a utility function which increases as its achieved timely throughput increases.
\end{itemize}

To the best of our knowledge, this is the first presented study on these three
fundamental problems centering around timely throughput for delay-constrained input-queued switches.
We should emphasize that delay-constrained input-queued switches are completely different from delay-unconstrained ones.
In delay-unconstrained scenarios, since packets will never expire and can be kept in the queues forever, the arrival traffic pattern
does not make a big difference (actually only the arrival rate matters in the capacity region characterization and in the throughput-optimal scheduling
policy design \cite{dai2000throughput}). However, in delay-constrained scenarios, since packets will expire if they are not scheduled
before their deadlines, the arrival traffic pattern has a significant impact on timely throughput. Thus, as compared to delay-unconstrained ones,
there are new challenges to study delay-constrained input-queued switches.

In this work, as a first step toward answering the above
three fundamental problems for delay-constrained input-queued switches,
we mainly study a special traffic pattern, called \emph{frame-synchronized traffic pattern}.
Such a traffic pattern can find applications in CPSs \cite{kim2012cyber}.
It was also the first focus in delay-constrained wireless communication \cite{hou2009qos,hou2010utility,hou2010utility,deng2017timely}. We also discuss how to consider more general traffic patterns.
In this work, we use three different approaches to study the above three fundamental problems. The three approaches come from different angles
and all have their own merits.
We summarize the results in Table \ref{tab:result-summary} and detail them as follows:
\begin{itemize}
\item The first approach is based on Markov Decision Process (MDP) theory. MDP has a strong modeling capability. Since our system is Markovian (though deterministic),
we can use MDP to model our problem. By leveraging results in \cite{deng2017timely}, in Sec.~\ref{sec:MDP},
we characterize the capacity region, design a feasibility-optimal scheduling policy, and  design a network-utility-maximization scheduling policy.
Due to its strong modeling capability, the MDP-based approach can be extended to more general traffic pattern, similar to \cite{deng2017timely}.
However, the MDP approach suffers from \emph{the curse of dimensionality}: it has an exponential complexity with the switch size.
\item The second approach exploits the problem's combinatorial features. By leveraging some results in combinatorial matrix theory,
in Sec.~\ref{sec:capacity-region}, we characterize the capacity region with only a polynomial number of linear constraints (see \eqref{equ:capacity-region}).
This breaks the curse of dimensionality of the first MDP-based approach for capacity region characterization.
\item The third approach is based on the framework of Lyapunov optimization.
By leveraging the Lyapunov-drift theorem \cite{neely2010stochastic}, in Sec.~\ref{sec:Lyapunov}, we show that the problem of minimizing Lyapunov drift  is
a maximum-weight $T$-disjoint-matching problem. We further design a polynomial-time algorithm to optimally solve  the maximum-weight $T$-disjoint-matching problem
based on the bipartite-graph edge-coloring algorithm. We show that our maximum-weight $T$-disjoint-matching scheduling policy (called $T$-MWM)
is feasibility-optimal.
\end{itemize}

We remark that although it is straightforward to apply the MDP-based approach in  \cite{deng2017timely}
to solve our three fundamental problems, the solutions are of exponential complexity and thus cannot be efficiently
applied to large-size switches.
Therefore, the polynomial-time capacity region characterization in \eqref{equ:capacity-region} and the polynomial-time feasibility-optimal  $T$-MWM
scheduling policy are two main contributions of this paper. These two results also serve as the delay-constrained counterparts of the capacity region characterization
and the throughput-optimal maximum-weight-matching scheduling policy for the delay-unconstrained input-queued switch
in \cite{mckeown1999achieving}.

\emph{Notation.} In this paper, we define set $[C] \triangleq \{1,2,\cdots, C\}$ for any positive integer $C$.
We use calligraphy font to denote sets, e.g., $\mathcal{A}$.
We use bold math font to denote vectors and matrices whose entries use the corresponding normal font, e.g., $\bm{b}=(b_{t}:  t \in [T]), \bm{R}=(R_{i,j}:  i,j \in [N])$.
We sometimes omit the index range of vectors/matrices if it is not ambiguous in the context, e.g.,  $\bm{b}=(b_{t}), \bm{R}=(R_{i,j})$.
We use upper-case letter to denote random variables, e.g., $S$.

\section{System Model and Problem Formulations}

\subsection{System Model} \label{subsec:model}
\begin{figure}
  \centering
  \subfigure[Input-queued switch]{
    \label{fig:switch-diagram} 
    \includegraphics[width=0.725\linewidth]{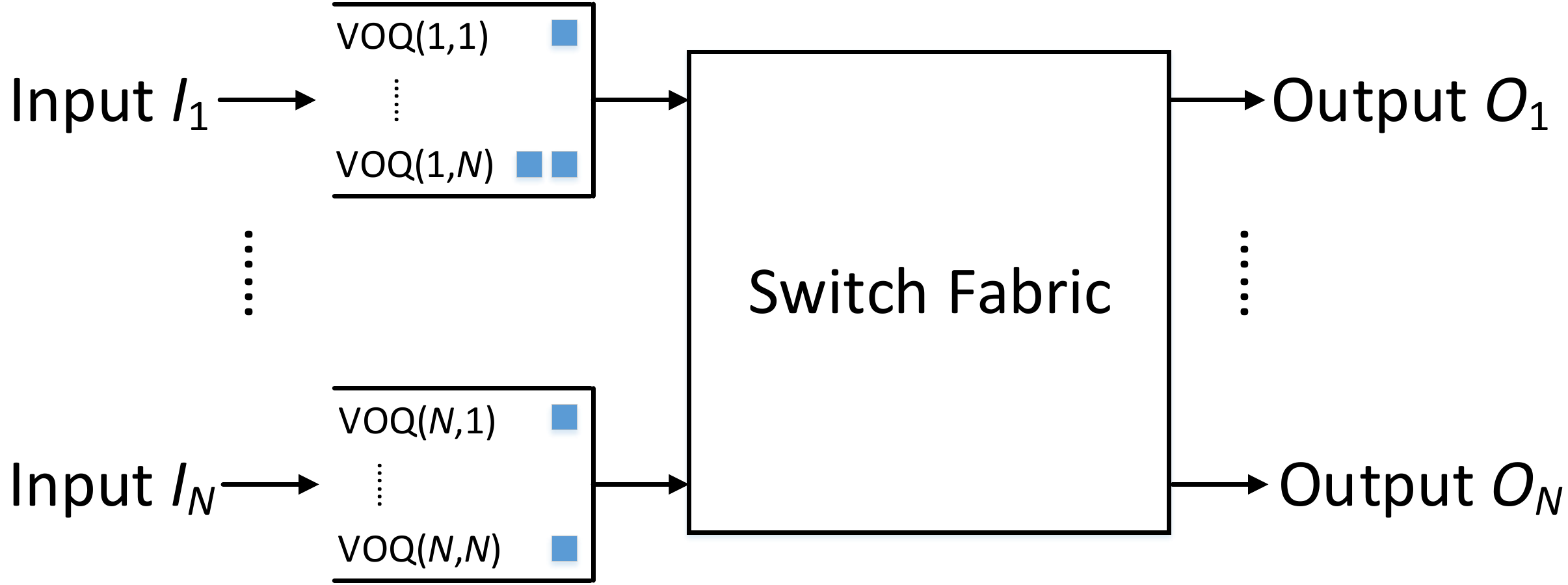}}
    \hfill
    \subfigure[Bipartite graph]{
    \label{fig:bipartite-graph} 
    \includegraphics[width=0.228\linewidth]{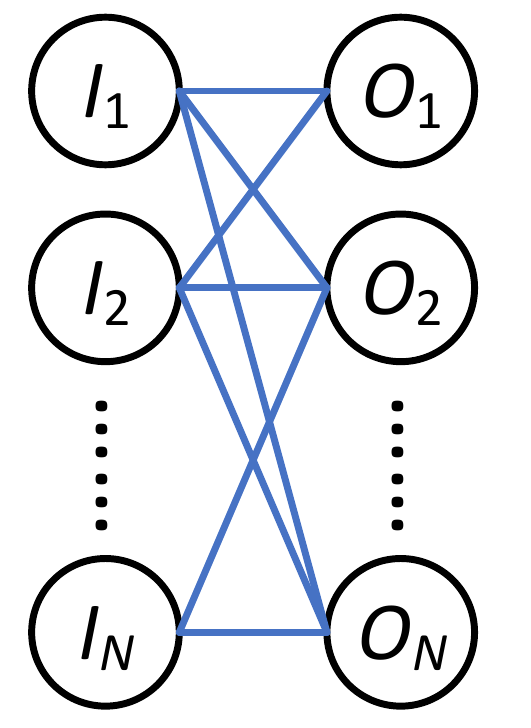}}
  \caption{An $N \times N$ input-queued switch using virtual output queueing (VOQ) and its corresponding bipartite graph $\mathcal{G} = (\mathcal{V}, \mathcal{E})$.} \vspace{-0.3cm}
  \label{fig:switch} 
\end{figure}

\textbf{Input-Queued Switch.}
We consider an $N\times N$ input-queued switch using virtual output queueing as shown in Fig.~\ref{fig:switch-diagram}.
Each input $I_i$ has $N$ virtual output queues (VOQs), denoted as VOQ$(i,j), \forall j \in [N] $.
VOQ($i,j$) contains all packets from input $I_i$ to output $O_j$.

\begin{figure}[t]
  \centering
  \includegraphics[width=0.8\linewidth]{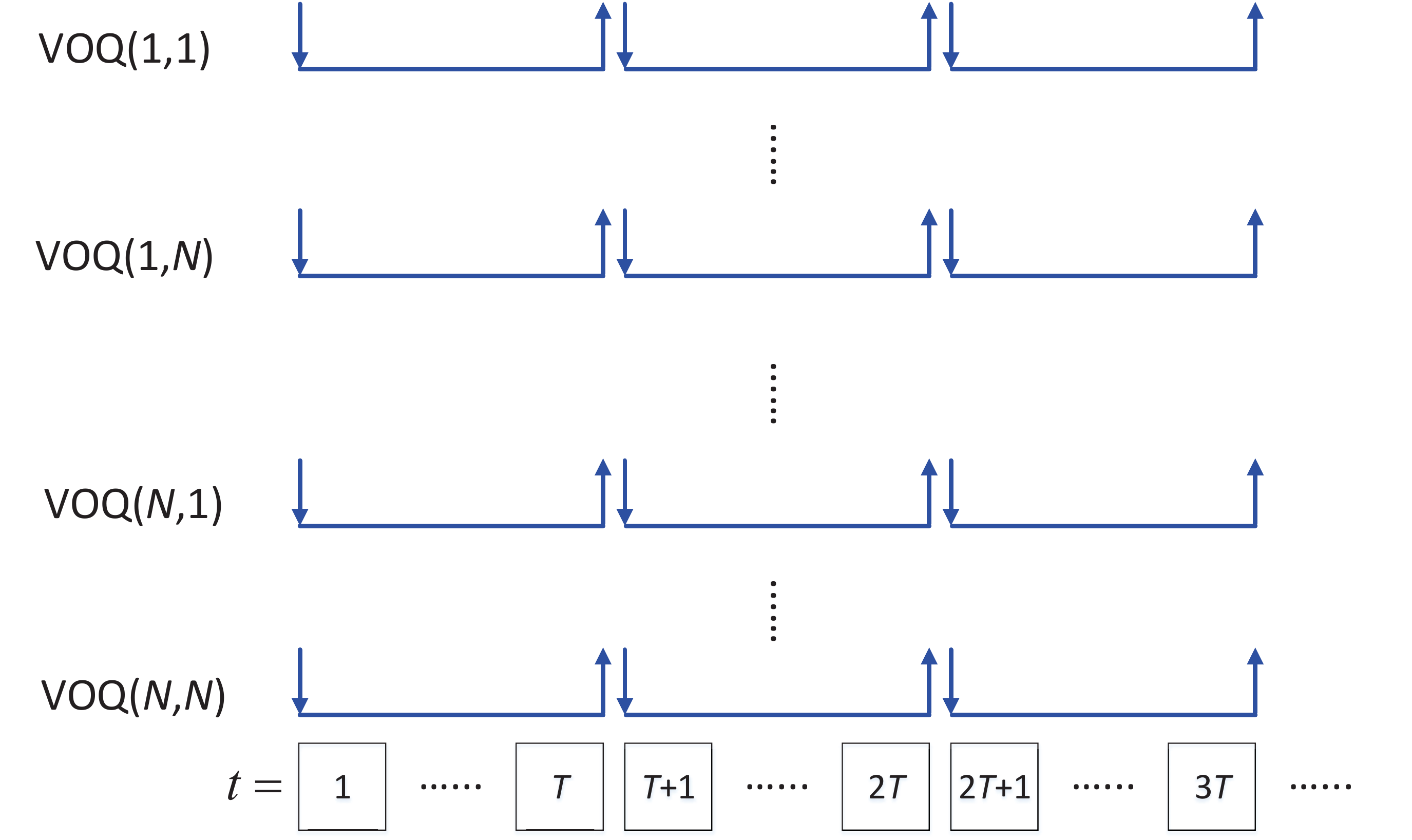}\\
  \caption{The frame-synchronized traffic pattern for the input-queued switch.} \vspace{-0.2cm}
  \label{fig:traffic}
\end{figure}

\textbf{Traffic Pattern.}
We consider a time-slotted system. We assume a \emph{frame-synchronized} traffic pattern \cite{hou2009qos}:
starting from slot 1, there is an incoming packet for each VOQ every $T$ slots and the deadline of any packet is also $T$ slots.
We call $T$ the frame length. Such a traffic pattern is shown in Fig.~\ref{fig:traffic}. If a packet is delivered before its deadline, it
contributes to the throughput; otherwise, the packet is useless and will be dropped/discarded from the system.

The frame-synchronized traffic pattern can find applications in  CPSs \cite{kim2012cyber}.
In addition, like the delay-constrained wireless communication community \cite{hou2009qos,hou2010utility,kang2016performance,deng2017timely},
the special frame-synchronized traffic pattern is a good starting point  to investigate delay-constrained input-queued switches.
We also show that our first approach (the MDP-based approach)
can be extended to  more general traffic patterns in Sec.~\ref{sec:MDP}.

\textbf{Scheduling Algorithm/Policy.}
In each slot, the switch fabric can transmit some packets from the inputs to the outputs.
In this paper, we use the most common crossbar switch fabric.
However, due to the physical limitations of crossbar switch fabric,
each input can transmit at most one packet per slot and each output can receive at most one packet per slot. This is also known as
the crossbar constraints \cite{neely2007logarithmic}.
The crossbar switch is non-blocking in the sense that all packets satisfying the crossbar constraints
can be routed simultaneously in a slot.
For the $N \times N$ input-queued switch, we can construct a corresponding bipartite graph $\mathcal{G} = (\mathcal{V}, \mathcal{E})$
between the $N$ inputs and the $N$ outputs
where $\mathcal{V}= \{I_1,I_2, \cdots, I_N\} \cup \{O_1, O_2, \cdots, O_N\}$
and $\mathcal{E}=\{(I_i, O_j): i,j \in [N]\}$, as shown in Fig.~\ref{fig:bipartite-graph}.
Then the (deterministic) decision in each slot corresponds to
a matching\footnote{Recall that a matching in a graph is a set of pairwise non-adjacent edges; namely, no two edges share a common vertex.}
in the bipartite graph $\mathcal{G}$. More specifically, we denote a matching as a matrix $\bm{M}=(M_{i,j}: i,j \in [N])$,
where $M_{i,j}=1$ if edge $(I_i, O_j)$ is in the matching (i.e., VOQ($i,j$) is selected) and $M_{i,j}=0$ otherwise. Clearly, matching $\bm{M}$
should satisfy\footnote{With a little bit abuse of notation,
here we refer matrix $\bm{M}$ as the edge set in this matching and thus we call it matching $\bm{M}$.}
\vspace{-0.3cm}
\bse
\label{equ:matching}
\bee
& \sum_{j=1}^{N} M_{i,j} \le 1, \forall i \in [N], \label{equ:matching-i} \\
& \sum_{i=1}^{N} M_{i,j} \le 1, \forall j \in [N], \label{equ:matching-j} \\
& M_{i,j} \in \{0, 1\}, \forall i,j \in [N], \label{equ:matching-0-1}
\eee
\ese
where \eqref{equ:matching-i} restricts that any input $I_i$ can at most transmit one packet to one output and \eqref{equ:matching-j}
restricts that any output $O_j$ can at most receive one packet from one input.
We denote the set of all matchings as $\mathcal{M}$, i.e.,
\be
\mathcal{M} \triangleq \{\bm{M}=(M_{i,j}: i,j \in [N]):  \bm{M} \text{ satisfies } \eqref{equ:matching-i}-\eqref{equ:matching-0-1}\}. \nnb
\ee
The decision could also be randomized in  that it could randomly choose a matching among multiple matchings.
A scheduling algorithm/policy is the set of (possibly randomized) decisions at all slots.
We give two definitions for later analysis.
\begin{definition} \label{def:disjoint-matching}
Two matchings $\bm{M}=(M_{i,j})$ and $\bm{M}'=(M'_{i,j})$ are \emph{disjoint} if there does not exist a position $(i,j)$ such that
both $M_{i,j}=1$ and $M'_{i,j}=1$.
\end{definition}

\begin{definition} \label{def:T-disjoint-matching}
If $\bm{M}^t=(M^t_{i,j})$ is a matching for any $t \in [T]$,
we call the collection $\{\bm{M}^t: t \in [T]\}$ a \emph{$T$-disjoint matching}
if any two of them are disjoint.
\end{definition}

\subsection{Problem Formulations} \label{subsec:formulation}
For a scheduling policy $\pi$, we define the \emph{timely throughput} \cite{hou2009qos, deng2017timely}
from input $I_i$ to output $O_j$ as\footnote{We also call it the timely throughput of VOQ($i,j$).}
\be
R^{\pi}_{i,j} \triangleq \liminf_{ t \to \infty} \frac{\mathbb{E}\left[ \sum_{\tau=1}^{t} D^{\pi}_{i,j,\tau}\right]}{t}, \forall i,j \in [N],
\label{equ:R-i-j}
\ee
where $D^{\pi}_{i,j, \tau}=1$ if a packet is delivered from input $I_i$ to output $O_j$ at slot $\tau$ under scheduling policy $\pi$ and $D^{\pi}_{i,j,\tau}=0$ otherwise.
Here the expectation is taken over the randomness of matchings
if randomized matchings are specified in the scheduling policy $\pi$.
Since all expired packets will be removed from the system,
the timely throughput $R^{\pi}_{i,j}$ is the per-slot average number of delivered packets before expiration for VOQ$(i,j)$.
Note that we allow packet dropout/expiration and thus do not need to deliver all traffic packets. However, packet dropout/expiration
affects the timely throughput.

A rate matrix $\bm{R} = (R_{i,j})$ is \emph{feasible} if there exists a scheduling policy
such that the timely throughput from input $I_i$ to output $O_j$ is at least $R_{i,j}$ for all $i,j  \in [N]$.
We then define the \emph{capacity region} $\mathcal{R}(T)$ as the set of all feasible rate matrices with frame length $T$.

Based on these definitions, in this paper, we study the following three  timely-throughput-centric fundamental problems:
\begin{itemize}
\item How to characterize the capacity region $\mathcal{R}(T)$?
\item How to design a feasibility-optimal scheduling policy, i.e.,
to design a policy that can support any feasible rate matrix $\bm{R} \in \mathcal{R}(T)$?
\item How to design a scheduling policy to maximize
the network utility, i.e.,
\be \label{equ:problem-num}
\max_{\bm{R} \in \mathcal{R}(T)} \sum_{i=1}^N \sum_{j=1}^N U_{i,j}(R_{i,j}),
\ee
where each input-output pair $(i,j)$ has an increasing, concave, and continuously differentiable utility function $U_{i,j}(R_{i,j})$ with respect to its achieved timely throughput $R_{i,j}$?
\end{itemize}
The capacity region problem is important because it serves as the foundation to evaluate any scheduling policy.
The feasibility-optimal scheduling policy design problem is important for inelastic delay-constrained applications which have
stringent minimum timely throughput requirements.
The network-utility-maximization scheduling policy design
problem is important for elastic delay-constrained applications which do not have stringent minimum timely throughput requirements
but obtain larger utility for larger timely throughput. Next we propose three different approaches to solve the above three fundamental problems
for delay-constrained input-queued switches.

\section{An MDP-based Approach} \label{sec:MDP}
For delay-constrained wireless communication, the authors in \cite{deng2017timely} proposed
a unified MDP-based formulation to study three fundamental problems similar to ours. By observing our system
is also Markovian (though deterministic), we can also use MDP theory \cite{puterman2014markov} to solve our three fundamental problems.
\ifx \ISTR \undefined
Since it is similar to apply the MDP-based approach of \cite{deng2017timely} to our problems,
we present the details in our technical report \cite{TR},
where we show that in principle the MDP-based approach solves all three fundamental problems in Sec.~\ref{subsec:formulation}.
\else
Our MDP can be described by a tuple $\{\mathcal{S}, \mathcal{A}, \{P_t\}, \{{r}_{i,j}\}\}$,
where $\mathcal{S}$ is the state space, $\mathcal{A}$ is the action space, $P_t(\bm{s}'|\bm{s},\bm{a})$ is the transition probability
from state $\bm{s}$ to state $\bm{s}'$ if taking action $\bm{a}$ at slot $t$, and ${r}_{i,j}(\bm{s},\bm{a})$ is the per-slot reward of VOQ($i,j$) if the state
is $\bm{s}$ and the action is $\bm{a}$.

\textbf{State.}
For VOQ$(i,j)$, we define its state at slot $t$ as
\be
S_{i,j,t}=\left\{
  \begin{array}{ll}
    1, & \hbox{if there exists a packet in VOQ$(i,j)$ at slot $t$;} \\
    0, & \hbox{otherwise.}
  \end{array}
\right.
\ee
The system state at slot $t$ is denoted as
\be
\bm{S}_t = (S_{i,j,t}: i,j \in [N]).
\ee
Then the state space $\mathcal{S}$ is the set of all $\{0,1\}$ $N \times N$ matrices.\footnote{Recall that a matrix is a $\{0,1\}$ matrix
if all its entries are either 0 or 1.} The total number of states is $|\mathcal{S}|=2^{N^2}$.

\textbf{Action.}
Let us define
\be
A_{i,j,t}=\left\{
  \begin{array}{ll}
    1, & \hbox{if VOQ$(i,j)$ is selected at slot $t$;} \\
    0, & \hbox{otherwise.}
  \end{array}
\right.
\ee
Then the action at slot $t$ is denoted as
\be
\bm{A}_t = (A_{i,j,t}: i,j \in [N]).
\ee
Due to the crossbar constraints, our action at slot $t$ must be a matching of $\mathcal{G}$, i.e., $\bm{A}_t \in \mathcal{M}$.
In our MDP formulation, we further restrict our action at each slot to be a \emph{perfect matching} without loss of optimality.
A perfect matching is a matching such that any input/output is incident to an edge in the matching. Namely, if $\bm{M} \in \mathcal{M}$ is
a perfect matching, then all inequalities in \eqref{equ:matching-i} and \eqref{equ:matching-j} hold as equalities.
The reason that we can restrict our action to be a perfect matching without loss of optimality is as follows: for our bipartite graph $\mathcal{G}$ with $N$ inputs and $N$ outputs, if a matching $\bm{M} \in \mathcal{M}$ is non-perfect, then at least one input and at least one output
are not incident to the edges in matching $\bm{M}$ and thus we can add a new edge to construct a new matching.
We can keep adding new edges to finally construct a perfect matching $\bm{M}'$. Since $\bm{M}'$ is a superset of $\bm{M}$,
any VOQ selected in action $\bm{M}$ will also be selected in action $\bm{M}'$. Thus, it suffices to consider perfect matchings.
Then the action space $\mathcal{A}$ is the set of all perfect matchings. For our $N\times N$ input-queued switch, we have in total $|\mathcal{A}|=N!$ perfect matchings.

\textbf{Transition Probability.}
For our input-queued switch, we have a deterministic transition which depends on three events: (i) packet expiration, (ii) packet arrival, and (iii) packet delivery.
For VOQ$(i,j)$, the transition probability is as follows:
\begin{itemize}
\item When $t \neq fT, f \in \mathbb{Z}^+$, since there is no packet expiration and no packet arrival, $\forall s_{i,j} \in \{0,1\}$, we have
\[
P_{i,j,t}(s'_{i,j}|s_{i,j}, a_{i,j}=1)=
\left\{
  \begin{array}{ll}
    1, & \hbox{if $s'_{i,j}=0$;} \\
    0, & \hbox{if $s'_{i,j}=1$;}
  \end{array}
\right.
\]
\[
P_{i,j,t}(s'_{i,j}|s_{i,j}, a_{i,j}=0)=
\left\{
  \begin{array}{ll}
    1, & \hbox{if $s'_{i,j}=s_{i,j}$;} \\
    0, & \hbox{if $s'_{i,j}\neq s_{i,j}$.}
  \end{array}
\right.
\]
\item
When $t = fT$ for some $f \in \mathbb{Z}^+$, since the old packet (if any) will expire and a new packet will arrive, $\forall s_{i,j} \in \{0,1\}, a_{i,j} \in \{0,1\}$,
we have
\[
P_{i,j,t}(s'_{i,j}|s_{i,j}, a_{i,j})=
\left\{
  \begin{array}{ll}
    1, & \hbox{if $s'_{i,j}=1$;} \\
    0, & \hbox{if $s'_{i,j}=0$.}
  \end{array}
\right.
\]
\end{itemize}

Then by the fact that given action $\bm{a}=(a_{i,j}: i,j \in [N])$ at slot $t$,
the transition of each individual state $s_{i,j}$ is independent of all other states,
the system transition from state $\bm{s}=(s_{i,j}: i,j \in [N])$ to state $\bm{s}'=(s'_{i,j}: i,j \in [N])$
if taking action $\bm{a}=(a_{i,j}: i,j \in [N])$ at slot $t$ is
\be
P_t(\bm{s}'|\bm{s},\bm{a})=\Pi_{i=1}^N \Pi_{j=1}^N P_{i,j,t}(s'_{i,j}|s_{i,j}, a_{i,j}).
\ee
Note that the transition probability is not stationary because it depends on slot $t$
(specifically, it depends on whether $t\neq fT$ or $t=fT$).

\textbf{Reward.}
The per-slot reward of VOQ$(i,j)$ under state $\bm{s}=(s_{i,j}: i,j \in [N])$  and action $\bm{a}=(a_{i,j}: i,j \in [N])$ is
\be
r_{i,j}(\bm{s},\bm{a}) =
\left\{
  \begin{array}{ll}
    1, & \hbox{if $s_{i,j}=1, a_{i,j}=1$;} \\
    0, & \hbox{otherwise.}
  \end{array}
\right.
\ee

After we formulate the MDP for our input-queued switch, we can easily see that the average reward (in the $\liminf$ sense)
of VOQ$(i,j)$ for a given scheduling policy in the MDP is the timely throughput of VOQ$(i,j)$
for the same scheduling policy in our system. Thus, we can solve the formulated
 MDP to answer the three problems in Sec.~\ref{subsec:formulation}.
Similar to \cite{deng2017timely}. Let $x_t(\bm{s}, \bm{a})$ be the joint probability that the system is in state $\bm{s}$ and the
action is $\bm{a}$ at slot $t$ and we then have the following result.

\begin{theorem} \label{thm:MDP}
(i) The network utility maximization problem in \eqref{equ:problem-num} can be solved by
the following linear-constrained convex optimization problem:
\bse \label{equ:convex}
\bee
\max & \quad \sum_{i=1}^N \sum_{j=1}^N U_{i,j}(R_{i,j}) \label{equ:convex-obj} \\
\text{s.t.} & \quad \sum_{\bm{a} \in \mathcal{A}} x_{t+1}(\bm{s}', \bm{a}) = \sum_{\bm{s} \in \mathcal{S}} \sum_{\bm{a} \in \mathcal{A}} P_t(\bm{s}'|\bm{s},\bm{a}) x_t(\bm{s},\bm{a}) \nnb \\
& \qquad \qquad  \forall \bm{s}' \in \mathcal{S}, t \in [T-1] \label{equ:convex-con1} \\
& \quad \sum_{a \in \mathcal{A}} x_{1}(\bm{s}',\bm{a}) = \sum_{\bm{s} \in \mathcal{S}} \sum_{\bm{a} \in \mathcal{A}} P_T(\bm{s}'|\bm{s}, \bm{a}) x_T(\bm{s},\bm{a}) \nnb \\
& \qquad \qquad  \forall \bm{s}' \in \mathcal{S} \label{equ:convex-con2} \\
& \quad R_{i,j} \le \frac{\sum_{t=1}^{T} \sum_{\bm{s} \in \mathcal{S}} \sum_{\bm{a} \in \mathcal{A}} {r}_{i,j}(\bm{s},\bm{a})x_t(\bm{s},\bm{a})}{T} \nnb \\
& \qquad \qquad \forall i,j \in [N] \label{equ:convex-con3} \\
& \quad \sum_{\bm{s} \in \mathcal{S}} \sum_{\bm{a} \in \mathcal{A}} x_t(\bm{s},\bm{a}) = 1, \quad \forall t \in [T] \label{equ:convex-con4} \\
\text{var.} & \quad x_t(\bm{s},\bm{a}) \ge 0, \quad \forall t \in [T], \bm{s} \in \mathcal{S}, \bm{a} \in \mathcal{A} \label{equ:convex-var1} \\
& \quad  R_{i,j} \ge 0, \quad \forall i,j \in [N]. \label{equ:convex-var2}
\eee
\ese

(ii) The capacity region $\mathcal{R}(T)$ can be characterized by
\bee \label{equ:capacity-regin-MDP}
& \mathcal{R}(T) = \left\{(R_{i,j}: i,j \in [N]): \text{There exists }\right.\nnb\\
&\left.\text{ a $\{x_t(\bm{s},\bm{a}): t \in [T], \bm{s} \in \mathcal{S}, \bm{a} \in \mathcal{A} \}$  }  \right.\nnb \\
& \left. \text{ such that \eqref{equ:convex-con1}-\eqref{equ:convex-var2} holds for $(R_{i,j}: i,j \in [N])$ }  \right\}.
\eee

(iii) For any $\{x_t(\bm{s},\bm{a})\}$ and  $(R_{i,j})$ satisfying \eqref{equ:convex-con1}-\eqref{equ:convex-var2},
the following randomized cyclo-stationary (RCS) policy achieves timely throughput $R_{i,j}$ for any VOQ$(i,j)$,
\be
\left\{
  \begin{array}{ll}
    \prob_{\bm{A}_t|\bm{S}_t}(\bm{a}|\bm{s})=\frac{ x_t(\bm{s},\bm{a})}{\sum_{\bm{a}' \in\mathcal{A}}x_t(\bm{s},\bm{a}')}, & \hbox{$\forall t \in [T]$;} \\
    \prob_{\bm{A}_t|\bm{S}_t}(\bm{a}|\bm{s})= \prob_{\bm{A}_{t-T}|\bm{S}_{t-T}}(\bm{a}|\bm{s}), & \hbox{$\forall t > T$.}
  \end{array}
\right.
\label{equ:RCS}
\ee
\end{theorem}

In \eqref{equ:convex}, since $x_t(\bm{s}, \bm{a})$ is the joint probability that the system is in state $\bm{s}$ and the
action is $\bm{a}$ at slot $t$,  \eqref{equ:convex-con1} and \eqref{equ:convex-con2}
are the consistency condition (or probability flow balance equation) for slots $1, 2, \cdots T-1$ and slot $T$, respectively.
Note that in \eqref{equ:convex-con2}, we come back to slot 1 from slot $T$ because we consider a frame of $T$ slots.
The right-hand side of \eqref{equ:convex-con3} is the per-slot average reward.
Note that in \eqref{equ:convex-con3} we use inequality because if we can support a  timely throughput $R_{i,j}$, we can always
support a smaller  timely throughput $R'_{i,j} < R_{i,j}$ without affecting other VOQs' timely throughput. However, due to the increasing property of the utility function $U_{i,j}(R_{i,j})$,
an optimal solution is always achieved with equality in \eqref{equ:convex-con3}.
Eq. \eqref{equ:convex-con4} says that the sum of probabilities over all states and all actions is 1.

Part (i) of Theorem \ref{thm:MDP} shows that we can get the optimal network utility by solving a linear-constrained convex optimization problem;
based on the optimal solution, part (iii) of Theorem \ref{thm:MDP} shows that we can construct a RCS policy to achieve such an optimal network utility.
Namely, we obtain a network-utility-maximization scheduling policy.
Part (ii) of Theorem \ref{thm:MDP} shows that the capacity region can be characterized by a finite (though exponentially increasing with respect to $N$) number of linear constraints in \eqref{equ:convex}.
Then for any feasible timely throughput matrix $\bm{R} \in \mathcal{R}(T)$, we can input it in \eqref{equ:convex} as a set of given variables and
then after solving problem \eqref{equ:convex} (with any valid utility functions),
we get a feasible solution; based on this feasible solution, part (iii) of Theorem \ref{thm:MDP} again shows that
we can construct a RCS policy to achieve the given feasible timely throughput matrix $\bm{R} \in \mathcal{R}(T)$.
Namely, we obtain a feasibility-optimal scheduling policy.
Therefore, Theorem~\ref{thm:MDP} shows that in principle our MDP-based approach solves all three fundamental problems in Sec.~\ref{subsec:formulation}.
\fi

In addition, although we mainly study the frame-synchronized traffic
pattern, we should further remark that our MDP-based approach  can also be extended
to more general traffic patterns which might be non-framed
or non-synchronized and could have stochastic arrivals. This
is similar to  \cite{deng2017timely}, which extends the frame-synchronized traffic
pattern to general traffic patterns for delay-constrained wireless
communication problems.

However, MDP framework suffers from the curse of dimensionality:
the number of states is $2^{N^2}$ and the number of actions is $N!$, both
increasing exponentially with respect to the switch size $N$.
\ifx \ISTR \undefined
Specifically, the MDP-based capacity region characterization has $O(T \cdot 2^{N^2} \cdot N!)$ linear equalities/inequalities
and the per-frame time complexity of the MDP-based scheduling policy is $O(N!)$.
\else
Specifically, the MDP-based capacity region characterization has $O(T \cdot 2^{N^2} \cdot N!)$ linear equalities/inequalities.
For the MDP-based RCS scheduling policy, we need a table of size $O(T \cdot 2^{N^2} \cdot N!)$
to store the solution $\{x_t(\bm{s},a)\}$ and then based on an observed state $\bm{s}$ in a frame,
the per-frame time complexity to obtain the action distribution $\prob_{\bm{A}_t|\bm{S}_t}(\bm{a}|\bm{s})$
is $O(N!)$.
\fi
To break the curse of dimensionality,
next in Sec.~\ref{sec:capacity-region}, we exploit the combinatorial features of our problem which are hidden by our MDP-based approach
and give a new capacity region characterization with only a polynomial number of linear constraints; and
in Sec.~\ref{sec:Lyapunov}, we propose a polynomial-time feasibility-optimal scheduling policy.

\section{A Simple Capacity Region Characterization} \label{sec:capacity-region}
In this section, by exploiting the combinatorial features of the problem,
we give a simple capacity region characterization in terms of only a polynomial number of linear constraints for the delay-constrained input-queued switches.
Toward that end, we first present some preliminary definitions and results.
\begin{definition}[\cite{brualdi1991combinatorial}]
An $N\times N$ square matrix $\bm{E} = (E_{i,j})$ is doubly substochastic  if it satisfies the following conditions:
\be \label{equ:def-dss}
\left\{
  \begin{array}{ll}
    \sum_{j=1}^{n} E_{i,j} \le 1, & \hbox{$\forall i \in [N]$;} \\
    \sum_{i=1}^{n} E_{i,j} \le 1, & \hbox{$\forall j \in [N]$;} \\
    E_{i,j} \ge 0, & \hbox{$\forall i, j \in [N]$.}
  \end{array}
\right.
\ee
\end{definition}

Denote $\mathcal{J}$ as the set of all doubly substochastic $N \times N$ matrices and let $k$ be a positive integer.
Let $\mathcal{J}_k$ be the set of all $1/k$-bounded doubly substochastic $N \times N$ matrices, i.e.,
\be
\mathcal{J}_k \triangleq \{\bm{E} \in \mathcal{J}: E_{i,j} \in [0, 1/k], \forall i,j \in [N]\}.
\ee
Denote $\mathcal{H}_k$ as the set of all matrices in $\mathcal{J}_k$ whose entries are either $0$ or $1/k$. Clearly, $\mathcal{H}_k$
is a finite set. We now give a convex-hull characterization for set $\mathcal{J}_k$.

\begin{lemma}[\protect{\cite[Theorem 1]{deng2017convex}}]
\label{lem:dss-convex-hull}
$\mathcal{J}_k$ is the convex hull of all matrices in $\mathcal{H}_k$.
\end{lemma}

Lemma \ref{lem:dss-convex-hull} shows  that any matrix $\bm{E} \in \mathcal{J}_k$ can be expressed
as a convex combination of some matrices in  $\mathcal{H}_k$.

A matrix is a subpermutation matrix if it is a $\{0,1\}$ matrix and  each of its line (row or column) has at most one 1.
It is straightforward to see that matrix $\bm{M}$ is a matching (i.e., $\bm{M} \in \mathcal{M}$)
if and only if $\bm{M}$ is a subpermutation matrix.
In addition, a matrix is a $k$-subpermutation matrix for some positive integer $k$ if
it is a $\{0,1\}$ matrix and the sum of each line (row or column) is at most $k$.
We give a decomposition result for $k$-subpermutation matrices.

\begin{lemma}[\protect{\cite[Theorem 4.4.3]{brualdi1991combinatorial}}]
\label{lem:T-subpermutation-decomposition}
Any $k$-subpermutation matrix can be expressed as the sum of $k$ subpermutation matrices.\footnote{In the original result \cite[Theorem 4.4.3]{brualdi1991combinatorial},
the maximum line sum of the matrix is exactly $k$. However, if the maximum line sum of the matrix is less than $k$,
\cite[Theorem 4.4.3]{brualdi1991combinatorial} shows that we can decompose it into less than $k$ subpermutation matrices.
We can further add some zero matrices such that we can decompose it into exactly $k$ subpermutation matrices.
}
\end{lemma}

With the help of the above-mentioned results, we now give a new capacity region characterization for our delay-constrained input-queued switch.

\begin{theorem} \label{the:capacity-region}
The capacity region $\mathcal{R}(T)$ is the set of all rate matrices $\bm{R}=(R_{i,j})$ satisfying the following linear inequalities:
\bse \label{equ:capacity-region}
\bee
& \sum_{i=1}^N R_{i,j} \le 1, \forall j \in [N], \label{equ:capacity-region-con1} \\
& \sum_{j=1}^N R_{i,j} \le 1, \forall i \in [N], \label{equ:capacity-region-con2} \\
& R_{i,j} \in [0, {1}/{T}], \forall i,j \in [N]. \label{equ:capacity-region-con3}
\eee
\ese
\end{theorem}
\begin{IEEEproof}
The necessity of this result can be easily proved.
Since any output $O_j$ can at most receive one packet per slot, the aggregate timely throughput involving output $O_j$ is at most 1 and thus
\eqref{equ:capacity-region-con1} holds;
since any input $I_i$ can at most transmit one packet per slot, the aggregate timely throughput involving input $I_i$
is at most 1 and thus \eqref{equ:capacity-region-con2} holds;
since every VOQ has only one packet in a frame of $T$ slots, its (per-slot) timely throughput is at most $1/T$
and thus  \eqref{equ:capacity-region-con3} holds.
Thus, any feasible rate matrix $\bm{R}$ must satisfy \eqref{equ:capacity-region}.
Then we only need to show that any rate matrix $\bm{R}$ satisfying \eqref{equ:capacity-region} can be achieved by some scheduling policy.

Clearly any matrix $\bm{R}$ satisfying \eqref{equ:capacity-region} is a $1/T$-bounded doubly substochastic matrix. Then from Lemma \ref{lem:dss-convex-hull},
we know that $\bm{R}$ can be expressed as a convex combination of a finite number of (say in total $K$) doubly substochastic matrices  whose  entries are either 0 or $1/T$, i.e.,
\be \label{equ:R-decomposition}
\bm{R} = \sum_{k=1}^{K} \lambda_k \bm{R}_k,
\ee
where $\lambda_k > 0, \sum_{k=1}^{K} \lambda_k = 1$ and matrix $\bm{R}_k$ is a doubly substochastic matrix with entries being 0 or $1/T$.
Since $\bm{R}_k$ is a doubly substochastic matrix, it has at most $T$ entries being $1/T$ in each line (row or column).

We multiply matrix $\bm{R}_k$ by $T$ and obtain matrix $T\bm{R}_k$. Clearly, the entry of matrix $T\bm{R}_k$
is either 0 or 1 and the sum of each line (row or column) is at most $T$, implying that $T\bm{R}_k$ is a $T$-subpermutation matrix.
Now according to Lemma \ref{lem:T-subpermutation-decomposition},  matrix $T\bm{R}_k$
can be decomposed as the sum of $T$ subpermutation matrices, i.e.,
\be \label{equ:TR-k-decomposition}
T\bm{R}_k = \sum_{t=1}^{T} \bm{M}_{k,t},
\ee
where $\bm{M}_{k,t}$ is a subpermutation matrix, which corresponds to a
matching. In addition, since the entry of matrix $T\bm{R}_k$ is either 0 or 1,
all subpermutation matrices (matchings) $\bm{M}_{k,t}$'s are pairwise disjoint (see Definition~\ref{def:disjoint-matching}),
implying that $\{\bm{M}_{k,t}: t \in [T]\}$ is a $T$-disjoint-matching (see Definition~\ref{def:T-disjoint-matching}).

Combining \eqref{equ:R-decomposition} and \eqref{equ:TR-k-decomposition}, we have
\be \label{equ:TR-decomposition}
T\bm{R} = \sum_{k=1}^{K} \lambda_k \cdot T\bm{R}_k = \sum_{k=1}^{K} \lambda_k \sum_{t=1}^{T} \bm{M}_{k,t}.
\ee

Then we construct the following scheduling policy: in each frame, select the $T$-disjoint-matching $\{\bm{M}_{k,t}: t \in [T]\}$
with probability $\lambda_k$ for any $k \in [K]$.
Here when we select the $T$-disjoint-matching   $\{\bm{M}_{k,t}: t \in [T]\}$ in any frame $f=0,1,2,\cdots$, we do the scheduling as follows:
\begin{itemize}
\item Perform matching $\bm{M}_{k,1}$ at slot $fT+1$;
\item Perform matching $\bm{M}_{k,2}$ at slot $fT+2$;
\item $\cdots$
\item Perform matching $\bm{M}_{k,T}$ at slot $(f+1)T$.
\end{itemize}

For the $T$-disjoint-matching   $\{\bm{M}_{k,t}: t \in [T]\}$,
if $$\left(\sum_{t=1}^{T} \bm{M}_{k,t}\right)_{i,j} = (T\bm{R}_k)_{i,j} = 1,$$  VOQ$(i,j)$ will be scheduled in a frame; otherwise,
VOQ$(i,j)$ will not be scheduled. Since we select all (in total $K$)  $T$-disjoint-matchings randomly according to probability distribution $\{\lambda_k\}$,
the probability to schedule VOQ$(i,j)$ (which is also the expected number of delivered packets for VOQ$(i,j)$) in a frame is
\be
\sum_{k=1}^{K} \lambda_k (T\bm{R}_k)_{i,j} = (T\bm{R})_{i,j} = T R_{i,j},
\ee
where the first equality follows from \eqref{equ:TR-decomposition}.
Therefore, the (per-slot) timely throughput of VOQ$(i,j)$ is $\frac{T {R}_{i,j}}{T}= {R}_{i,j}$ for any VOQ$(i,j)$.
This completes the proof.
\end{IEEEproof}

Theorem \ref{the:capacity-region} gives a new capacity region characterization \eqref{equ:capacity-region}
with only $2N^2+2N=O(N^2)$ linear inequalities, much lower than the exponential-size MDP-based characterization (which needs $O(T \cdot 2^{N^2} \cdot N!)$ linear equalities/inequalities).
We further make some remarks for Theorem \ref{the:capacity-region}.

\subsection{Comparison with Delay-Unconstrained Results}
Our capacity region characterization for delay-constrained input-queued switches
has a similar non-overbooking condition (see \eqref{equ:capacity-region-con1}, \eqref{equ:capacity-region-con2}) with that for delay-unconstrained ones \cite{mckeown1999achieving,dai2000throughput,chang1999service,chang2000birkhoff,chang2002load},
except that each VOQ's timely throughput is upper bounded by $1/T$ (see \eqref{equ:capacity-region-con3}).
However, there is a fundamental difference --- in delay-unconstrained input-queued switches, the capacity region
is in terms of the (incoming) arrival rate of all VOQs,
while in our delay-constrained ones, the capacity region is in terms of the (achieved) timely throughput of VOQs.
In other words, we allow packet loss/expiration and characterize the fundamental limit of  timely throughput in our delay-constrained input-queued switches.
\ifx \ISTR \undefined
\rev{We also compare our proof technique for Theorem \ref{the:capacity-region} with that for the capacity region characterization
for delay-unconstrained input-queued switches \cite{chang1999service,chang2000birkhoff} in our technical report \cite{TR},
}
\else
In addition,
we also cannot directly apply the proof techniques for the capacity region characterization based on Birkhoff-von Neumann decomposition approach
for delay-unconstrained input-queued switches \cite{chang1999service,chang2000birkhoff}.
In \cite{chang1999service,chang2000birkhoff}, the proof of the capacity region characterization for delay-unconstrained input-queued switches
relies on two results:
\begin{itemize}
\item (1) For any $N \times N$ doubly substochastic matrix $\bm{R}$, there exists an $N \times N$
doubly stochastic matrix $\bm{R}'$ (where the first two inequalities in \eqref{equ:def-dss}
hold as equalities) such that entry-wise $\bm{R} \le \bm{R}'$. This was proved by John von Neumann in \cite{von1953certain};
\item (2) Any  $N \times N$ doubly stochastic matrix can be expressed as the convex combination of some $N \times N$ permutation matrices.
This was proved by Garrett Birkhoff in \cite{birkhoff}.
\end{itemize}

If we followed this proof technique to prove our Theorem~\ref{the:capacity-region}, we need the following two results:
\begin{itemize}
\item (1') For any $N \times N$ $1/T$-bounded doubly substochastic matrix $\bm{R}$, there exists an $N \times N$ $1/T$-bounded
doubly stochastic matrix $\bm{R}'$ such that entry-wise $\bm{R} \le \bm{R}'$;
\item (2') Any $N \times N$ $1/T$-bounded doubly stochastic matrix can be expressed as the convex combination of some $N \times N$
doubly stochastic matrices  whose entries are either 0 or $1/T$.
\end{itemize}
Part (2') holds according to \cite{watkins1974convex}. However,
it turns out that part (1') does \emph{not} hold. As a counter-example, consider the following
$N \times N$ $1/T$-bounded doubly substochastic matrix $\bm{R}$ where $N=3$, $T=2$, and
\[
\bm{R}=
\left(
  \begin{array}{ccc}
    0.5 & 0.4 & 0.1 \\
    0.4 & 0.5 & 0.1 \\
    0.1 & 0.1 & 0.1 \\
  \end{array}
\right).
\]
If we need to find an $N \times N$ $1/T$-bounded
doubly stochastic matrix $\bm{R}'$ such that entry-wise $\bm{R} \le \bm{R}'$,
we can only increase $R_{3,3}$, but we can at most increase up to $R_{3,3}=0.5$ and the resulting matrix is still
not a doubly stochastic matrix. Hence, there does not exist an $N \times N$ $1/T$-bounded
doubly stochastic matrix $\bm{R}'$ such that entry-wise $\bm{R} \le \bm{R}'$ and thus
part (1') does {not} hold.
Therefore, we cannot apply the same proof idea of delay-unconstrained input-queued switches
into our delay-constrained ones.

Instead, we have to leverage the result in Lemma~\ref{lem:dss-convex-hull} which directly gives a convex-hull characterization
for $1/T$-bounded doubly substochastic matrices. This is a key step to prove Theorem~\ref{the:capacity-region}.

\fi

\subsection{The Special Case of $T \ge N$} \label{subsubsec:T-ge-N}
If $T \ge N$, we can see that the rate matrix $\bm{R} = (R_{i,j}=1/T: i,j \in [N])$ is
in the capacity region \eqref{equ:capacity-region},
which achieves the largest timely throughput for all VOQs.
In fact, \eqref{equ:capacity-region-con3} implies \eqref{equ:capacity-region-con1} and \eqref{equ:capacity-region-con2} when $T \ge N$.
Indeed, when $T \ge N$, we can construct a scheduling policy to transmit all packets without any packet loss/expiration
so as to attain a timely throughput of $1/T$ for all VOQs.
\ifx \ISTR \undefined
\rev{Please see the details of how to construct the scheduling policy in our technical report \cite{TR}.}
\else
We first note that a perfect matching can also be represented as a permutation of elements $\{1,2,\cdots,N\}$.
For example with $N=3$, permutation $(2,3,1)$ means the perfect matching $\{(I_1, O_2), (I_2, O_3), (I_3, O_1)\}$. Then in any frame $f=0,1,2,\cdots$, we do the following scheduling:
\begin{itemize}
\item Perform permutation  $(1,2,\cdots,N-1,N)$ at slot $fT+1$;
\item Perform permutation $(N,1,2,\cdots,N-1)$ at slot $fT+2$;
\item $\cdots$
\item Perform permutation $(2,\cdots, N-1, N, 1)$ at slot $fT+N$.
\item Perform nothing from slot $fT+N+1$ to slot $(f+1)T$.
\end{itemize}
Namely, starting with permutation $(1,2,\cdots,N-1,N)$, we keep doing right-circular shift for the obtained permutation,
which is similar to the idea of the ``circular-shift" matrix in \cite{chang2006providing}.
We can check that any VOQ is scheduled once (and only once) in the first $N$ slots of any frame.
Thus, all $N^2$ packets in any frame are delivered, implying that all packets in the system can be delivered.
\fi

\subsection{Lack of Scheduling Policy}
Note that when we prove the achievability part in Theorem~\ref{the:capacity-region},
we construct a randomized scheduling policy
based on the distribution $\{\lambda_k\}$. Although we show the existence of parameters $\{\lambda_k\}$,
we do \emph{not} know how to find such  $\{\lambda_k\}$. Thus, our constructed randomized scheduling policy
is only an existing policy but we do not have ways to implement it.

This is different from the result in delay-unconstrained Birkhoff-von Neumann input-queued switches
in \cite{chang1999service,chang2000birkhoff}, where
the authors utilized the fact that any $N \times N$ doubly stochastic matrix can be expressed as the convex combination of some $N \times N$ permutation matrices \cite{birkhoff},
and more importantly they proposed an algorithm of complexity $O(N^{4.5})$ to find the convex-combination parameters $\{\phi_k\}$.
Based on $\{\phi_k\}$, the authors in \cite{chang1999service,chang2000birkhoff} further implemented a throughput-optimal scheduling policy in polynomial time.

\section{A Polynomial-Time Feasibility-Optimal Scheduling Policy} \label{sec:Lyapunov}
The combinatorial approach in Sec.~\ref{sec:capacity-region} breaks the curse of dimensionality of
the MDP-based approach for the problem of characterizing the capacity region. In this section, we further
break the curse of dimensionality of the MDP-based approach for the problem of designing a feasibility-optimal scheduling policy.
In particular, we leverage the framework of Lyapunov optimization and design a polynomial-time feasibility-optimal scheduling policy
for our delay-constrained input-queued switches.

\begin{figure}[t]
  \centering
  \includegraphics[width=\linewidth]{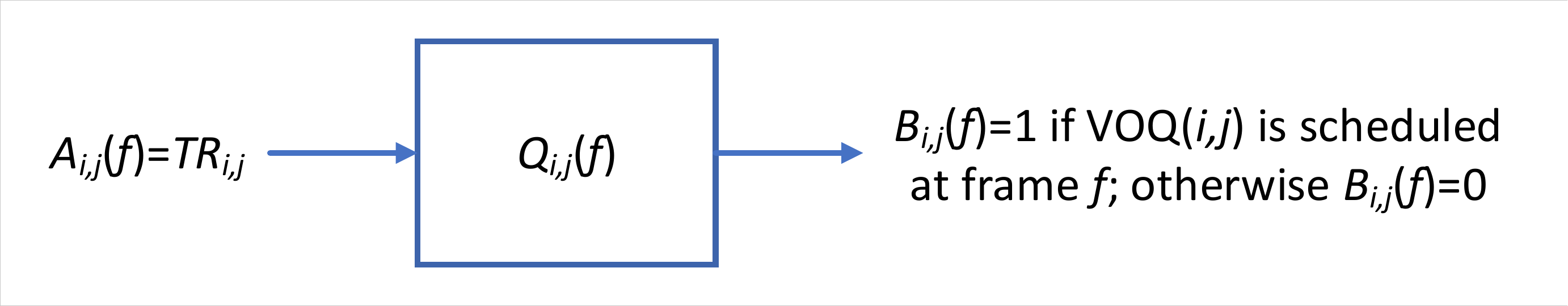}
  \caption{The constructed virtual queue.}\label{fig:virtual-queue}
\end{figure}

For any VOQ($i,j$), if it has a timely throughput requirement $R_{i,j}$,
we construct a virtual queue\footnote{Readers should distinguish virtual queue here from VOQ (virtual output queue).} as shown in Fig.~\ref{fig:virtual-queue}:
\begin{itemize}
\item The virtual queue is indexed by the frames in the real system, denoted as $f=0,1,2,\cdots$;
\item The arrival process of the virtual queue $A_{i,j}(f)$ is a constant flow with size $T R_{i,j}$ for any frame $f$;
\item The service process of the virtual queue $B_{i,j}(f)$ depends on the scheduling policy in the real system:
$B_{i,j}(f)= 1$ if VOQ$(i,j)$ is scheduled in frame $f$ in
the real system and $B_{i,j}(f)= 0$ otherwise;
\item By using the standard queue dynamics in \cite{neely2010stochastic}, the queue is updated as (with initial queue length $Q_{i,j}(0)=0$)
\be \label{equ:virtual-queue-update}
\resizebox{.88\linewidth}{!}{$Q_{i,j}(f+1) = \max\{Q_{i,j}(f)-B_{i,j}(f), 0\} + A_{i,j}(f).$}
\ee
\end{itemize}

Note that the virtual (queue) system is different from the real system. In the real system,
a packet expires at the end of its frame. However, in the virtual queue, all arrivals will not expire and always
stay in the virtual queue. Moreover, the time scale is also different: our virtual system is frame-based while our real system
is slot-based. We use $B_{i,j}(f)$ to connect the virtual system and real system.

According to the queue stability theorem \cite[Theorem 2.5(b)]{neely2010stochastic}, if the virtual queue $Q_{i,j}$ is mean rate stable, then
\be
\limsup_{F\rightarrow\infty} \frac{1}{F} \sum_{f=0}^{F-1} \mathbb{E}[A_{i,j}(f) - B_{i,j}(f)] \le 0. \label{equ:needly-theorem-2.5}
\ee
Since $A_{i,j}(f)=TR_{i,j}, \forall f$, then \eqref{equ:needly-theorem-2.5} implies,
\be
\liminf_{F\rightarrow\infty} \frac{1}{F} \sum_{f=0}^{F-1} \mathbb{E} [B_{i,j}(f)] \ge TR_{i,j}.
\ee
Note that $\liminf_{F \to \infty}(1/F)\sum_{f=0}^{F-1} \mathbb{E}[B_{i,j}(f)]$ is the achieved per-frame timely throughput for VOQ($i,j$) in the real
system. Hence the achieved (per-slot) timely throughput for VOQ($i,j$) in the real system is
 $$\liminf_{F \to \infty} \frac{1}{TF}\sum_{f=0}^{F-1} \mathbb{E}[B_{i,j}(f)] \ge \frac{T R_{i,j}}{T} = R_{i,j}.$$
Thus, to achieve timely throughput $R_{i,j}$ for VOQ$(i,j)$ is equivalent to make the virtual queue $Q_{i,j}$ mean rate stable.

By using the Lyapunov-drift theorem \cite[Theorem 4.1]{neely2010stochastic}, it is standard to show that
the following maximum-weight scheduling policy can make all virtual queues mean rate stable:
in each frame $f=0,1,2,\cdots$, select a matrix $\bm{B}(f) = \left(B_{i,j}(f): i,j \in [N]\right)$ (which corresponds to $T$ matchings in this frame of in total $T$ slots)
to maximize the queue weight sum, i.e.,
\be
\max_{\bm{B}(f)} \sum_{i=1}^N \sum_{j=1}^N Q_{i,j}(f) B_{i,j}(f). \label{equ:max-weight-problem}
\ee
Our frame-based maximum-weight problem \eqref{equ:max-weight-problem} is different from the slot-based one for delay-unconstrained input-queued switches \cite{mckeown1999achieving,dai2000throughput},
which is exactly the classical maximum-weight-matching problem.
In \eqref{equ:max-weight-problem}, we need to find $T$ matchings to solve the frame-based maximum-weight problem.
Recall that $B_{i,j}(f)=1$ if VOQ($i,j$) is selected in frame $f$.
Even if VOQ$(i,j)$ is scheduled more than once in frame $f$, $B_{i,j}(f)$ is still 1 and cannot increase the objective value in \eqref{equ:max-weight-problem}.
This implies that there is no need to schedule the same VOQ for more than once in a frame.
Thus, it suffices to find $T$ \emph{disjoint} matchings
to solve \eqref{equ:max-weight-problem}, i.e., to select $\bm{B}(f)$ in frame $f$ is equivalent to select a $T$-disjoint-matching (see Definition~\ref{def:T-disjoint-matching}) of the bipartite graph $\mathcal{G}$.

Since in each frame we need to solve the same problem \eqref{equ:max-weight-problem} (though with different queue lengths/weights),
let us ignore the frame index $f$.
The problem to find a $T$-disjoint-matching with virtual queue weights $(Q_{i,j})$
to maximize the queue weight sum can be formulated as an integer linear programming (ILP),
\vspace{-0.5cm}
\bse \label{equ:T-matching-ILP}
\bee
\max & \quad \sum_{t=1}^{T}\sum_{i=1}^N \sum_{j=1}^N Q_{i,j} b_{i,j}^{t}  \label{opt-pre}\\
\text{s.t.} & \quad \sum_{i=1}^{N} b_{i,j}^{t} \le 1, \forall j \in [N], t \in [T] \label{equ:T-matching-ILP-con1}  \\
& \quad \sum_{j=1}^{N} b_{i,j}^{t} \le 1, \forall i \in [N], t \in [T]  \label{equ:T-matching-ILP-con2} \\
& \quad \sum_{t=1}^{T} b_{i,j}^{t} \le 1, \forall i,j \in [N]  \label{equ:T-matching-ILP-con3} \\
\text{var.} & \quad b_{i,j}^{t} \in \{0,1\}, \forall i,j \in [N], t \in [T] \label{equ:T-matching-ILP-con4}
\eee
\ese
In \eqref{equ:T-matching-ILP}, constraints \eqref{equ:T-matching-ILP-con1} and \eqref{equ:T-matching-ILP-con2} restrict that
at any slot $t \in [T]$, we select a matching $\bm{b}^t=(b^t_{i,j}) \in \mathcal{M}$; constraint \eqref{equ:T-matching-ILP-con3}
restricts that all matchings selected in $T$ slots of the frame are pairwise disjoint, i.e., $\{\bm{b}^t: t \in [T]\}$ is a $T$-disjoint-matching.
Based on \eqref{equ:T-matching-ILP}, we can simply reconstruct $B_{i,j} = \sum_{t=1}^{T} b_{i,j}^{t}$ to solve problem \eqref{equ:max-weight-problem}.

A nature approach to solve ILP \eqref{equ:T-matching-ILP} is to iteratively apply the (per-slot) maximum-weight matching algorithm.
However, as we show in
\ifx \ISTR \undefined
our technical report \cite{TR},
\else
Appendix~\ref{app:greed-max-weight},
\fi
the greedy iterative maximum-weight-matching algorithm is strictly suboptimal to ILP \eqref{equ:T-matching-ILP}.
This indicates that it is nontrivial to solve ILP \eqref{equ:T-matching-ILP}.
To solve ILP \eqref{equ:T-matching-ILP} optimally and efficiently,
we establish equivalence between ILP \eqref{equ:T-matching-ILP} and the following new ILP:
\bse \label{equ:T-matching-ILP2}
\bee
\max & \quad \sum_{i=1}^N \sum_{j=1}^N Q_{i,j} c_{i,j} \label{opt1}\\
\text{s.t.} & \quad \sum_{i=1}^{N} c_{i,j} \le T, \forall j \in [N] \label{equ:T-matching-ILP2-con1} \\
& \quad \sum_{j=1}^{N} c_{i,j} \le T, \forall i \in [N] \label{equ:T-matching-ILP2-con2} \\
\text{var.} & \quad c_{i,j} \in \{0,1\}, \forall i,j \in [N]
\eee
\ese
In \eqref{equ:T-matching-ILP2}, we find a set of VOQs to maximize the sum of their queue length/weight such that each input/outout is incident to at most $T$ VOQs.
From the perspective of bipartite graph $\mathcal{G}$,
ILP \eqref{equ:T-matching-ILP2} is to find a set of edges to maximize their weight sum such that each
node is incident to at most $T$ edges. We now establish the equivalence between ILP \eqref{equ:T-matching-ILP} and ILP \eqref{equ:T-matching-ILP2}.

\begin{theorem} \label{thm:ILP-is-equivalent-to-ILP2}
The optimal values of ILPs \eqref{equ:T-matching-ILP} and \eqref{equ:T-matching-ILP2} are the same.
Moreover, for any optimal solution $\{c_{i,j}\}$ to \eqref{equ:T-matching-ILP2}, we can use the bipartite-graph edge-coloring algorithm
to construct an optimal solution $\{b^t_{i,j}\}$ to \eqref{equ:T-matching-ILP} in polynomial time.
\end{theorem}
\begin{IEEEproof}
(i) For any feasible solution $\{b^t_{i,j}\}$ to ILP \eqref{equ:T-matching-ILP}, we construct
\[
c_{i,j} = \sum_{t=1}^{T} b^t_{i,j}.
\]
We can easily check that $\{c_{i,j}\}$ is feasible to ILP \eqref{equ:T-matching-ILP2} and the objective value of ILP \eqref{equ:T-matching-ILP2} is equal to that of  ILP \eqref{equ:T-matching-ILP} since
\[
 \sum_{i=1}^N \sum_{j=1}^N Q_{i,j} c_{i,j} =  \sum_{i=1}^N \sum_{j=1}^N Q_{i,j} \sum_{t=1}^{T} b^t_{i,j} = \sum_{t=1}^{T} \sum_{i=1}^N \sum_{j=1}^N  Q_{i,j}b^t_{i,j}.
\]
Therefore, the optimal value of  ILP \eqref{equ:T-matching-ILP2} is an upper bound of the optimal value of ILP \eqref{equ:T-matching-ILP}.

(ii) For any feasible solution $\{c_{i,j}\}$ to \eqref{equ:T-matching-ILP2}, we construct a bipartite graph $\mathcal{G}' =(\mathcal{V}, \mathcal{E}')$ where $(I_i, O_j) \in \mathcal{E}'$
if $c_{i,j}=1$. Due to \eqref{equ:T-matching-ILP2-con1} and \eqref{equ:T-matching-ILP2-con2}, we know that
the maximum degree of all nodes in $\mathcal{G}'$ is at most $T$.
The edge-coloring problem for a graph is to use minimum number of colors to color all edges such that any two edges sharing a common node do not have
the same color. It is well-known that the edges of
any bipartite graph can be colored with $\Delta$ colors \cite{konig1916graphen,schrijver1998bipartite,cole2001edge}
where $\Delta$ is the maximum node degree. 
Thus, our graph $\mathcal{G}'$ can be colored with at most $T$ colors.
Clearly, the set of all edges sharing the same color forms a matching of  bipartite graph $\mathcal{G}'$ and all
such (at most $T$) matchings are disjoint. For any matching, we can represent it as $(b_{i,j}: i,j \in [N])$ where $b_{i,j}=1$ if edge $(I_i,O_j)$ is in the matching.
We can also add some dummy/empty matchings such that we have in total $T$ disjoint matchings, i.e., constructing a
feasible solution $\{b^t_{i,j}:  i,j \in [N], t \in [T] \}$ to problem \eqref{equ:T-matching-ILP}. Since  all edges in graph $\mathcal{G}'$ is colored by those (at most $T$) colors,
we thus have
\[
\sum_{t=1}^{T} b^t_{i,j} = c_{i,j}, \forall i,j \in[N]
\]
and further
\[
\sum_{t=1}^{T} \sum_{i=1}^N \sum_{j=1}^N Q_{i,j}b^t_{i,j} =   \sum_{i=1}^N \sum_{j=1}^N Q_{i,j} \sum_{t=1}^{T} b^t_{i,j} = \sum_{i=1}^N \sum_{j=1}^N Q_{i,j} c_{i,j}
\]
implying that the objective value of  ILP \eqref{equ:T-matching-ILP} is equal to that of ILP  \eqref{equ:T-matching-ILP2}.
Thus the optimal value of  ILP \eqref{equ:T-matching-ILP2}  is a lower bound of the optimal value of ILP \eqref{equ:T-matching-ILP}.

Part (i) and part (ii) show that the optimal values of ILP \eqref{equ:T-matching-ILP} and ILP \eqref{equ:T-matching-ILP2} are the same.
Thus, the construction in part (ii) for any optimal solution to \eqref{equ:T-matching-ILP2} results in  an optimal solution to \eqref{equ:T-matching-ILP}.

It is well-known that we can color the bipartite graph $\mathcal{G}'$ with minimal number of (at most $T$) colors in polynomial time
\cite{konig1916graphen,schrijver1998bipartite,cole2001edge}. The best algorithm is that in \cite{cole2001edge} with complexity $O(N^2 \log T) = O(N^2 \log N)$.\footnote{Here we  require $T < N$.
Note that from the remark given in Sec.~\ref{subsubsec:T-ge-N}, we know that we can deliver all packets with a simple policy when $T \ge N$.
Thus, in this section, we only need to consider $T < N$.}
Therefore, once we obtain an optimal solution $\{c_{i,j}\}$ to \eqref{equ:T-matching-ILP2}, we can use the bipartite-graph edge-coloring algorithm
to construct an optimal solution $\{b^t_{i,j}\}$ to \eqref{equ:T-matching-ILP} in polynomial time.
\end{IEEEproof}

Theorem \ref{thm:ILP-is-equivalent-to-ILP2} shows that we only need to get an optimal solution to \eqref{equ:T-matching-ILP2}
 in order to get  an optimal solution $\{b^t_{i,j}\}$ to \eqref{equ:T-matching-ILP}.
Then the remaining problem is whether we can solve the new ILP  \eqref{equ:T-matching-ILP2} efficiently. Indeed,
problem \eqref{equ:T-matching-ILP2} can be solved in polynomial time.

\begin{algorithm}[t]
 \caption{The Maximum Weight $T$-Disjoint-Matching  Algorithm ($T$-MWM)}
 \label{alg:T-MWM}
\begin{algorithmic}[1]
    \REQUIRE A timely throughput matrix $\bm{R}=(R_{i,j}) \in \mathcal{R}(T)$
    \FOR{$i, j =1,2,\cdots, N$}
        \STATE Set $Q_{i,j}(0) = 0$
    \ENDFOR
    \FOR{$f=0, 1,2,\cdots$}
        \STATE Solve the relaxed LP of the vectorized version of ILP \eqref{equ:T-matching-ILP2} with weights $\{Q_{i,j}(f): i,j \in [N] \}$ by the simplex algorithm and get a vertex optimal solution $\{c_{i,j}\}$ \label{line:start}
        \STATE Construct a bipartite graph $\mathcal{G}'=(\mathcal{V}, \mathcal{E}')$ where $(I_i, O_j) \in \mathcal{E}'$ if $c_{i,j}=1$
        \STATE Use the bipartite-graph edge-coloring algorithm in \cite{cole2001edge} to color $\mathcal{G}'$ and  get a $T$-disjoint-matching  (possibly inserting some dummy/empty matchings)
              $\{(b^1_{i,j}),(b^2_{i,j}),\cdots,(b^T_{i,j})\}$ \label{line:end}
        \FOR{$t=1,2,\cdots, T$}
            \STATE Perform matching $(b^t_{i,j})$ at  slot $fT+t$
        \ENDFOR
        \FOR{$i=1,2,\cdots, N$}
            \FOR{$j=1,2,\cdots,N$}
                \STATE Set $B_{i,j}(f) = \sum_{t=1}^{T} b_{i,j}^t$
                \STATE Set $Q_{i,j}(f+1) = \max\{Q_{i,j}(f)-B_{i,j}(f), 0\} + TR_{i,j}$
        \ENDFOR
        \ENDFOR
    \ENDFOR
\end{algorithmic}
\end{algorithm}

It was shown in \cite{taylor2008optimal} that
the constraint matrix of the vectorized version of \eqref{equ:T-matching-ILP2} is totally unimodular.
Thus, we can resort to solving the relaxed LP of \eqref{equ:T-matching-ILP2} and any vertex optimal solution of the relaxed LP
would be integral and thus optimal to  \eqref{equ:T-matching-ILP2}. For example, the most widely-used simplex algorithm for LP
outputs a vertex optimal solution. Moreover, a recent result by Kitahara and Mizuno in \cite{kitahara2013bound}
shows that for an LP whose constraint matrix is totally unimodular and constraint constant vector is integral (which is indeed our case),
the number of different vertex solutions generated by the simplex method for this LP is polynomially bounded
by $n\lceil m ||b||_1 \log (m ||b||_1) \rceil$ where $n$ is the number of variables, $m$ is the number of constraints
and $b$ is the constraint constant vector. It is easy to see that $n=N^2$, $m=2N^2+2N$, and $||b||_1=2NT+N^2 \le 3N^2$
for our relaxed LP of \eqref{equ:T-matching-ILP2}. Thus, we can solve ILP  \eqref{equ:T-matching-ILP2} in polynomial time with complexity
$n\lceil m ||b||_1 \log (m ||b||_1) \rceil \le N^2\lceil (2N^2+2N) \cdot 3N^2 \cdot \log ((2N^2+2N) \cdot 3N^2)  \rceil = O(N^6 \log N)$.

Readers may wonder whether we can directly solve the relaxed LP of \eqref{equ:T-matching-ILP}, instead of leveraging an intermediate ILP \eqref{equ:T-matching-ILP2}.
It turns out that the direct approach does not work. We use an example
\ifx \ISTR \undefined
in our technical report \cite{TR}
\else
in Appendix~\ref{app:non-TU}
\fi
to show that the constraint matrix of the vectorized version of our original ILP
\eqref{equ:T-matching-ILP} is \emph{not} totally unimodular. Therefore, establishing the equivalence between ILP \eqref{equ:T-matching-ILP} and ILP \eqref{equ:T-matching-ILP2} is crucial.

We summarize the proposed maximum-weight $T$-disjoint-matching  scheduling algorithm in Algorithm~\ref{alg:T-MWM}, which we call $T$-MWM.
We now give a theorem to show that our $T$-MWM Algorithm is feasibility-optimal, i.e., it
 can achieve any feasible timely throughput requirements $\bm{R} \in \mathcal{R}(T)$.
\begin{corollary}
$T$-MWM is feasibility-optimal.
\end{corollary}
\begin{IEEEproof}
Based on Theorem~\ref{thm:ILP-is-equivalent-to-ILP2} and  \cite{taylor2008optimal},
we conclude that lines \ref{line:start}-\ref{line:end} of Algorithm~\ref{alg:T-MWM} ($T$-MWM) solves the
maximum-weight scheduling problem \eqref{equ:max-weight-problem}.
Thus, all virtual queues $(Q_{i,j})$ are mean rate stable according to the Lyapunov-drift theorem \cite[Theorem 4.1]{neely2010stochastic}.
Therefore, $T$-MWM is feasibility-optimal.
\end{IEEEproof}

Note that the per-frame time complexity of Algorithm~\ref{alg:T-MWM} ($T$-MWM) is polynomial
in the order of $O(N^6 \log N) + O(N^2 \log N) = O(N^6 \log N)$,
which is much faster than the exponential-time MDP-based policy (of order $O(N!)$).

\textbf{Remarks.} In this section, we adopt the Lyapunov-optimization framework to design a polynomial-time feasibility-optimal
scheduling policy. We should further remark that our virtual queue $V_{i,j}(f)$ defined in
\eqref{equ:virtual-queue-update} is also termed deficit in the delay-constrained wireless communication community
\cite{hou2009qos,deng2017timely,kang2016performance}. In particular, our maximum weight scheduling policy is similar to
the largest-deficit-first (LDF) scheduling policy. However, as compared with LDF scheduling policy which
only needs to select the flow with largest deficit in each slot, our maximum weight scheduling policy needs
to solve a more difficult combinatorial problem, i.e., ILP \eqref{equ:T-matching-ILP}. In addition to the capacity region characterization in Theorem~\ref{the:capacity-region},
our main contribution in this section is to show that ILP \eqref{equ:T-matching-ILP} is equivalent to another problem, i.e., ILP \eqref{equ:T-matching-ILP2},
which can be solved in polynomial time.

\section{Simulation}
In this section, we use simulation to evaluate our capacity region and scheduling policies.

First, we show that
the capacity region characterized by the MDP-based approach
\ifx \ISTR \undefined
(see the details in our technical report \cite{TR})
\else
in \eqref{equ:capacity-regin-MDP}
\fi
and the capacity region in \eqref{equ:capacity-region} characterized by the combinatorial approach are the same.
We simulate a $3 \times 3$ switch and vary the frame length $T$ from 1 to 5. Since it is difficult to visualize the
capacity region (of dimension $3 \times 3=9$), we solve the network-utility-maximization problem \eqref{equ:problem-num}
for two different capacity region characterizations.
We adopt a linear utility function $U_{i,j}(R_{i,j})=w_{i,j}R_{i,j}$
for each VOQ$(i,j)$. We randomly pick a weight matrix,
which is realized as
\[
\bm{w} = (w_{i,j}) =
\left(
  \begin{array}{ccc}
    0.70   &  0.84 & 0.54 \\
    0.51   &  0.92 & 0.44 \\
    0.10   &  0.30 & 0.28\\
  \end{array}
\right).
\]
Note that both the MDP-based approach and the combinatorial approach
characterize the capacity region in terms of some linear constraints.
Thus, under the linear utility functions,
the network-utility maximization problem \eqref{equ:problem-num} becomes
a linear programming (LP), whose constraints are different under two different capacity region characterizations.

\begin{figure}
  \centering
  \subfigure[]{
    \label{fig:capacity-region-verification} 
    \includegraphics[width=0.467\linewidth]{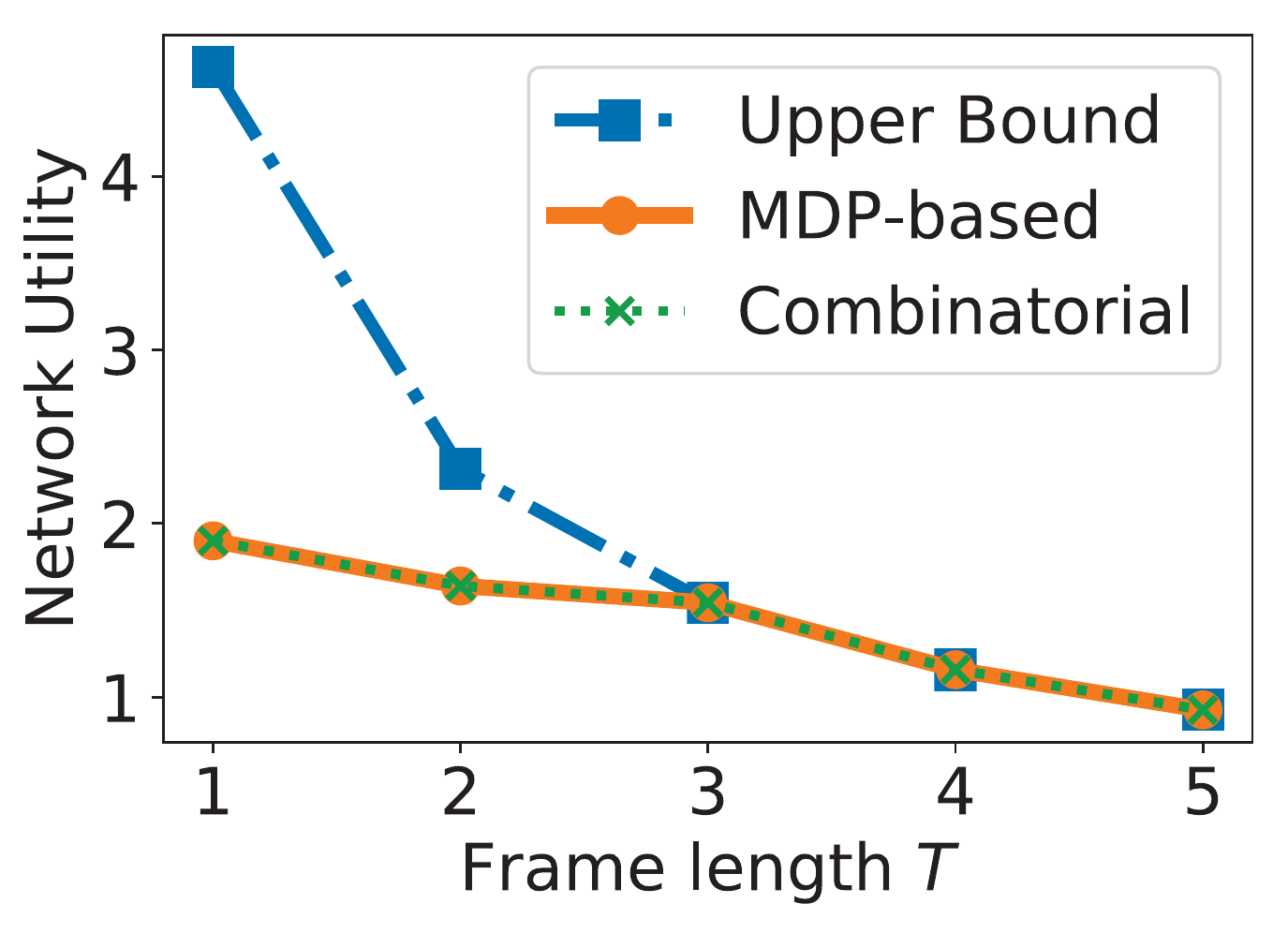}}
    \hfill
    \subfigure[]{
    \label{fig:scheduling-policies} 
    \includegraphics[width=0.493\linewidth]{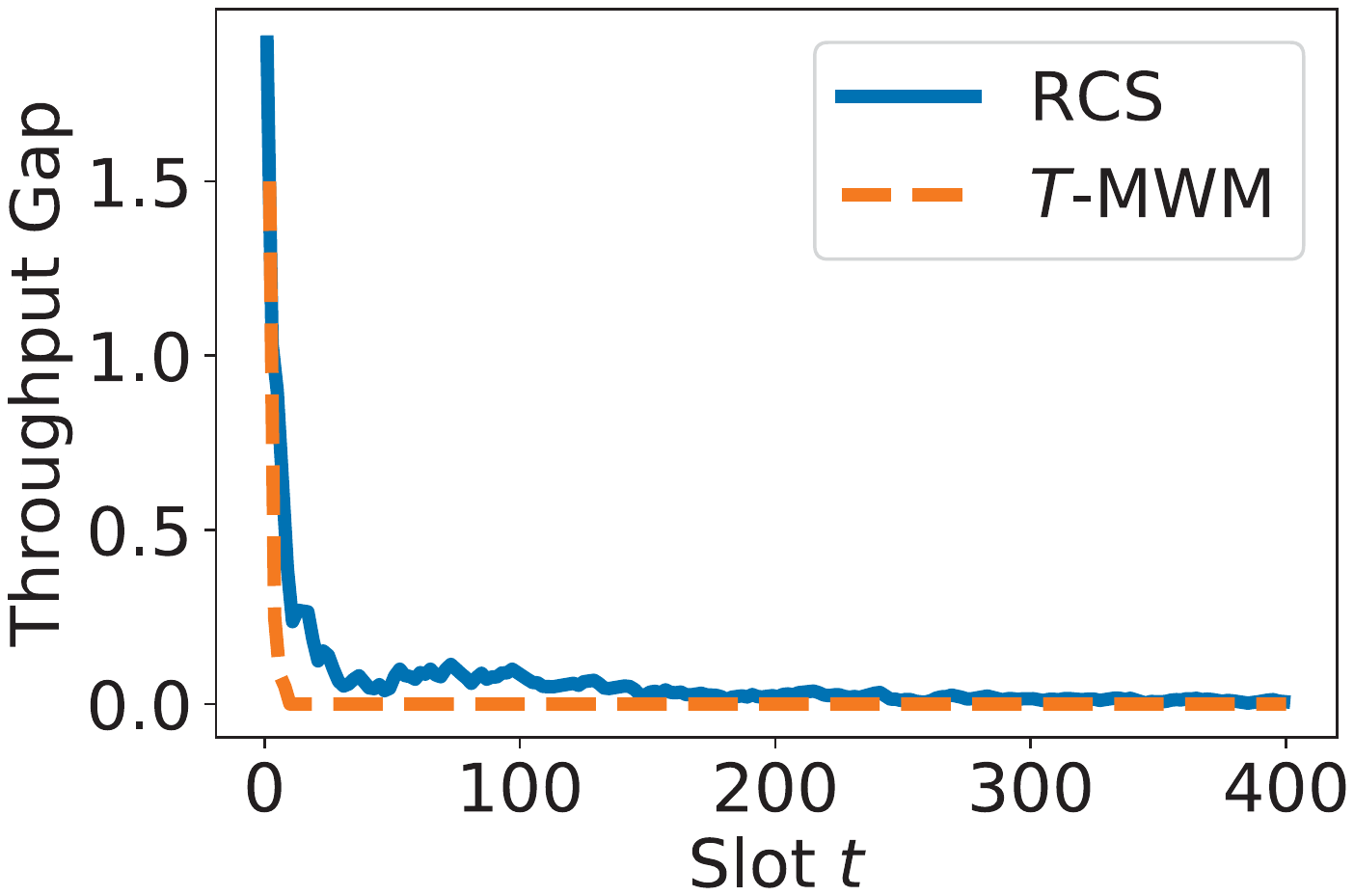}}
  \caption{Simulation for a $3 \times 3$ input-queued switch. (a)Verification of the equivalence   of   two  capacity region characterizations; (b) Evaluation for two  feasibility-optimal scheduling policies.} 
  \label{fig:simulation} 
\end{figure}

We show the achieved maximum network utility in Fig.~\ref{fig:capacity-region-verification}. We can see that
under two different capacity region characterizations, the achieved maximum network utilities are the same.
Namely, the two LPs with different linear constraints give the same optimal value.
We remark that such result holds for all our randomly generated weighted matrices,
verifying that our two different capacity region characterizations are the same.
In addition,  since each VOQ has only 1 packet every $T$ slots,
the  timely throughput of any VOQ is upper bounded by $1/T$ and  we thus plot the utility upper bound $\sum_{i=1}^N \sum_{j=1}^N w_{i,j}/T$
in Fig.~\ref{fig:capacity-region-verification}. We can see that indeed when $T \ge N=3$, the
achieved maximum network utility  attains the upper bound, verifying our discussion in Sec.~\ref{subsubsec:T-ge-N}.

Second, we compare our proposed two feasibility-optimal scheduling policies: the MDP-based algorithm (called RCS algorithm, \ifx \ISTR \undefined
see the details in our technical report \cite{TR})
\else
see \eqref{equ:RCS})
\fi
and $T$-MWM algorithm (Algorithm~\ref{alg:T-MWM}).
We again consider a $3 \times 3$  switch with $T=2$ and input a feasible rate matrix,
\[
\bm{R} = (R_{i,j}) =
\left(
  \begin{array}{ccc}
0.2 & 0.4  & 0.4 \\
0.3 & 0.5  & 0.2 \\
0.5 & 0.1  & 0.4 \\
  \end{array}
\right).
\]
We then run RCS and $T$-MWM. To verify that they are feasibility-optimal, we need to show that both can achieve
the target rate matrix  $\bm{R}$.
For any VOQ$(i,j)$, we obtain the empirical timely throughput up to slot $t$ as
\[
R^{\textsf{emp}}_{i,j}(t) \triangleq \frac{\sum_{\tau=1}^{t} D_{i,j,\tau}}{t},
\]
where $D_{i,j, \tau}=1$ if a packet is delivered from input $I_i$ to output $O_j$ at slot $\tau$ and $D_{i,j,\tau}=0$ otherwise.
We thus define the \emph{throughput gap} between the empirical rate matrix $\bm{R}^{\textsf{emp}}(t)$ and the target rate matrix $\bm{R}$ as
\be
\delta(\bm{R}^{\textsf{emp}}(t), \bm{R}) \triangleq \sum_{i=1}^N \sum_{j=1}^N \max\left\{R_{i,j} - R^{\textsf{emp}}_{i,j}(t), 0\right\}.
\label{def:throughput-gap}
\ee
Clearly, $\delta(\bm{R}^{\textsf{emp}}(t), \bm{R}) > 0$ if and only if there exists a VOQ which does not achieve its target timely throughput, i.e.,
$\exists i, j \in [N]$ such that $R^{\textsf{emp}}_{i,j}(t) < R_{i,j}$; and $\delta(\bm{R}^{\textsf{emp}}(t), \bm{R}) = 0$ if and only if every VOQ achieves the target timely throughput,
i.e., $R^{\textsf{emp}}_{i,j}(t) \ge R_{i,j}, \forall i,j \in [N]$. We show the throughput gap for all slots in Fig.~\ref{fig:scheduling-policies}. We can see
that the throughput gap converges to 0 in both algorithms, implying that both algorithms achieve the target rate matrix  $\bm{R}$. We remark that
such result holds for all our tried feasible rate matrices, verifying that both algorithms are feasibility-optimal.

Finally, we compare our proposed $T$-MWM scheduling policy with two baselines for the input-queued switch.
The first one is the (one-slot) maximum-weight-matching (MWM) scheduling policy that was proposed by \cite{mckeown1999achieving}
for delay-unconstrained input-queued switch.
MWM was proved to be throughput-optimal for delay-unconstrained traffic in \cite{mckeown1999achieving,dai2000throughput}.
The second one is the clearance-time-optimal (CTO) scheduling policy that was proposed by \cite{kang2013design} for real-time input-queued switch.
The authors in \cite{kang2013design} proved that CTO can minimize the maximum delivery delay among all packets (i.e., the clearance time).
Note that both MWM and CTO scheduling polices are not designed to route delay-constrained traffic where the hard deadline is
specified by different applications. Both MWM and CTO determine the schedule according to the length of real VOQs, while
our $T$-MWM determines the schedule according to the length of virtual queues \eqref{equ:virtual-queue-update}.

\begin{figure}
  \centering
  \subfigure[]{
    \label{fig:policy-compare-T} 
    \includegraphics[width=0.467\linewidth]{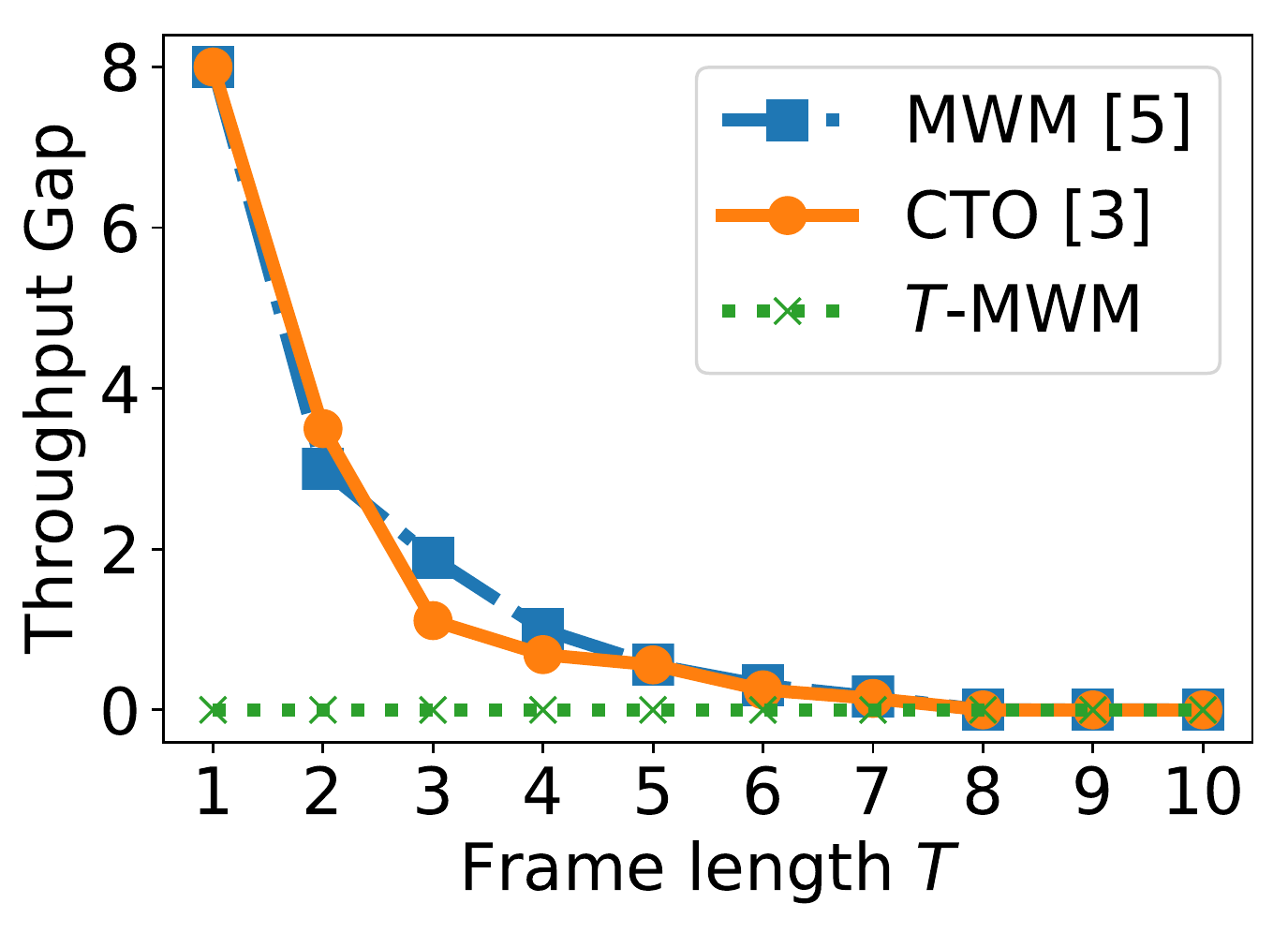}}
    \hfill
    \subfigure[]{
    \label{fig:policy-compare-N} 
    \includegraphics[width=0.493\linewidth]{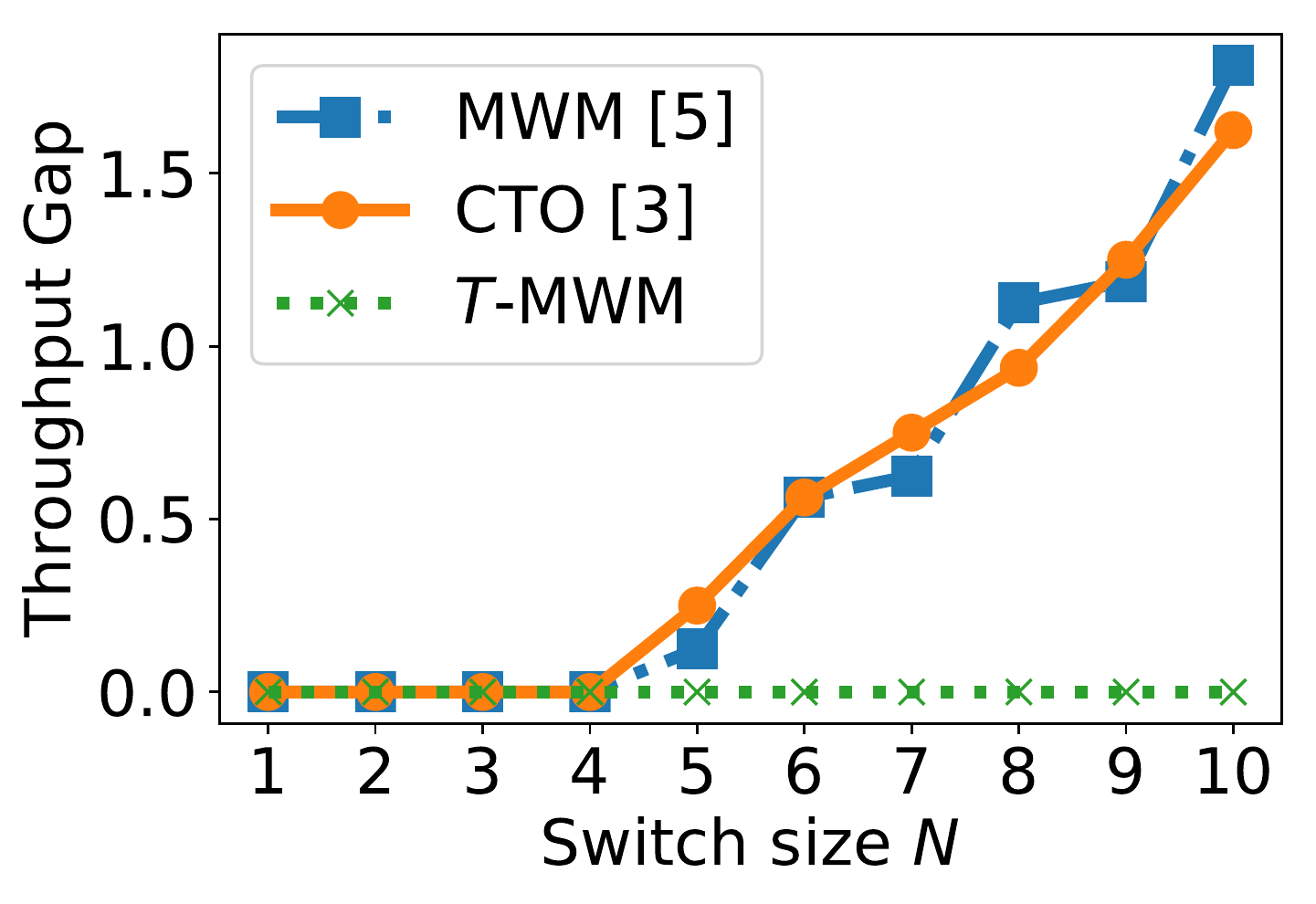}}
  \caption{Compare our proposed $T$-MWM policy with two baselines: (one-slot) maximum-weight-matching (MWM) scheduling policy that
  was proposed by \cite{mckeown1999achieving} for delay-unconstrained input-queued switch and the clearance-time-optimal (CTO) scheduling policy
  that was proposed by \cite{kang2013design} for real-time input-queued switch. (a) The switch size is $N=8$; (b) The frame length is $T=4$.} 
  \label{fig:policy-compare} 
\end{figure}

To compare these three scheduling policies in the delay-constrained setting,
for switch size $N$ and frame length $T$,
we randomly select  a weight matrix $\bm{w}$
and solve the network-utility-maximization problem $\max_{\bm{R} \in \mathcal{R}} \sum_{i,j\in[n]}w_{i,j}R_{i,j}$,
which gives us a feasible rate matrix $\bm{R}$. We then apply MWM, CTO, and $T$-MWM scheduling policies
to obtain the empirical timely throughput up to 10000 slots and finally we obtain the throughput gap based on \eqref{def:throughput-gap}.
We show the throughput gap of the three policies in Fig.~\ref{fig:policy-compare}, where
we fix the switch size to be $N=8$ and vary the frame length $T$ from 1 to 10 in Fig.~\ref{fig:policy-compare-T}
and we fix the frame length to be $T=4$ and vary the switch size $N$ from 1 to 10 in Fig.~\ref{fig:policy-compare-N}.
As we can see, our proposed $T$-MWM can achieve the target rate matrix $\bm{R}$ in any case, but neither
MWM nor CTO can achieve it when $N > T$. Thus, our proposed $T$-MWM policy outperforms both baselines
when the input-queued switch is required to deliver delay-constrained traffic.

\section{Conclusion}
To support delay-constrained traffic of real-time applications such as tactile Internet, networked control systems, and cyber-physical systems,
we study how to re-design the input-queued switch, which is the core component of communication networks.
We use three different approaches to solve the three fundamental problems
for delay-constrained input-queued switches centering around the performance metric of timely throughput.
The MDP-based approach can solve all three problems. In addition,
the MDP-based approach  can also be extended  to more general traffic patterns.
However, the MDP-based approach suffers from the curse of dimensionality.
To address this issue, we propose a combinatorial
approach to characterize the capacity region with only a polynomial number of linear constraints
and further propose a Lyapunov-based approach to design a polynomial-time feasibility-optimal
scheduling policy. In the future, it is  important to study how to design a polynomial-time network-utility-maximization scheduling policy,
how to efficiently extend to general traffic patterns to capture more practical scenarios, and how to implement our algorithms in practical switches.
In addition, it is interesting to study the system behaviour when we apply our algorithms to
the real communication system which deliver real-world delay-constrained traffic.

\ifx \ISTR \undefined
\else

\appendix

\section{Appendix}

\subsection{The Greedy Iterative Maximum-Weight Matching Algorithm is Not Optimal  to \eqref{equ:T-matching-ILP}} \label{app:greed-max-weight}
A nature approach to solve ILP \eqref{equ:T-matching-ILP}  is to iteratively apply the classical maximum-weight matching algorithm:
\begin{itemize}
\item Initialize an intermediate bipartite graph $\tilde{\mathcal{G}} = \mathcal{G} = (\mathcal{V}, \mathcal{E})$
and assign weight $Q_{i,j}$ to edge $(I_i, O_j)$ for any $(i,j)$;
\item Iterate $t$ from 1 to $T$;
\item In iteration $t$, we apply the classical maximum-weight matching for bipartite graph $\tilde{\mathcal{G}}$ and obtain
the corresponding solution $\bm{b}^t=(b^t_{i,j})$ where $\bm{b}^t$ is a matching.
For any $(i,j)$, remove edge $(I_i, O_j)$ in the bipartite graph $\tilde{\mathcal{G}}$ if $b^t_{i,j}=1$.
\end{itemize}
The final solution is a $T$-disjoint-matching  $\{\bm{b}^1, \bm{b}^2, \cdots, \bm{b}^T\}$.

This greedy algorithm is plausible because we iteratively strip out the maximum-weight matching.
However, it turns out that it may be strictly suboptimal. Let us see the following example.
Let $N=3, T=2$ and the queue weight matrix is
\[
\bm{Q}= (Q_{i,j})=
\left(
  \begin{array}{ccc}
    4 & 4 & 0 \\
    4 & 1 & 4 \\
    2 & 1 & 0 \\
  \end{array}
\right).
\]
Then the maximum weight sum is 17, which can be achieved by the following two disjoint matchings:
\[
\bm{b}^1= ({b}^1_{i,j}) =
\left(
  \begin{array}{ccc}
    1 & 0 & 0 \\
    0 & 0 & 1 \\
    0 & 1 & 0 \\
  \end{array}
\right), \quad
\bm{b}^2= ({b}^2_{i,j})  =
\left(
  \begin{array}{ccc}
    0 & 1 & 0 \\
    1 & 0 & 0 \\
    0 & 0 & 0 \\
  \end{array}
\right).
\]
However, when we apply the greedy algorithm, we get the following two disjoint matchings:
\[
\tilde{\bm{b}}^1= (\tilde{{b}}^1_{i,j}) =
\left(
  \begin{array}{ccc}
    0 & 1 & 0 \\
    0 & 0 & 1 \\
    1 & 0 & 0 \\
  \end{array}
\right), \quad
\tilde{\bm{b}}^2= (\tilde{{b}}^2_{i,j}) =
\left(
  \begin{array}{ccc}
    1 & 0 & 0 \\
    0 & 1 & 0 \\
    0 & 0 & 0 \\
  \end{array}
\right),
\]
which results in weight sum $10+5=15<17$.
Note that $\bm{b}^1$ is not a maximum-weight matching in iteration 1 but $\tilde{\bm{b}}^1$ is.
This example indicates that sometimes one should not
favor a maximum-weight matching in previous iterations
but instead leave some large weights into later matchings.

Thus, the greedy iterative maximum-weight matching algorithm may be strictly suboptimal.
To some extent, it also reveals the difficulty to solve ILP \eqref{equ:T-matching-ILP}.

\subsection{An Example to Show that the Constraint Matrix of the Vectorized Version of ILP \eqref{equ:T-matching-ILP} Is Not Totally Unimodular} \label{app:non-TU}

We first relax the binary variable to a real number and get the following linear programming (LP):
\bse \label{equ:T-matching-LP}
\bee
\max & \quad \sum_{t=1}^{T}\sum_{i,j \in [1,N]} Q_{i,j} b_{i,j}^{t}  \label{equ:T-matching-LP-obj}  \\
\text{s.t.} & \quad \sum_{i=1}^{N} b_{i,j}^{t} \le 1, \forall j\in[N], t\in[T]  \label{equ:T-matching-LP-con1} \\
& \quad \sum_{j=1}^{N} b_{i,j}^{t} \le 1, \forall i\in[N], t\in[T]  \label{equ:T-matching-LP-con2} \\
& \quad \sum_{t=1}^{T} b_{i,j}^{t} \le 1, \forall i,j\in[N]  \label{equ:T-matching-LP-con3} \\
\text{var.} & \quad b_{i,j}^{t} \ge 0, \forall i,j\in[N], t\in[T] \label{equ:T-matching-LP-con4}
\eee
\ese
We then vectorize the three-dimensional variables $(b_{i,j}^{t})$
according to dimension $j$, $i$ and $t$ in order and obtain a vector variable $\bm{b}$.
Then the constraint matrix in \eqref{equ:T-matching-LP} is

\be \label{equ:con-matrix-M}
\bm{C}=
\left(
  \begin{array}{cccc}
    \bm{L} & \bm{0} & \cdots & \bm{0} \\
    \bm{0} & \bm{L} & \cdots & \bm{0} \\
    \vdots & \vdots & \vdots & \vdots \\
    \bm{0} & \bm{0} & \cdots & \bm{L} \\
    \bm{I} & \bm{I} & \cdots & \bm{I}
  \end{array}
\right),
\ee
where $\bm{L}$ is the $2N\times N^2$ incident matrix of the bipartite graph $\mathcal{G}$,
$\bm{0}$ is $2N \times N^2$ zero matrix, and $\bm{I}$ is the $N^2 \times N^2$ identity matrix. In $\bm{C}$, we have $T$ repeats for $\bm{L}$ and $\bm{I}$.
The first $2NT$ rows in $\bm{C}$, i.e., the row of $\bm{L}$'s, correspond to constraints \eqref{equ:T-matching-LP-con1} and \eqref{equ:T-matching-LP-con2};
the second $N^2$ rows in $\bm{C}$, i.e., the row of $\bm{I}$'s, correspond to constraint \eqref{equ:T-matching-LP-con3}.

Then \eqref{equ:T-matching-LP} is vectorized as follows,
\be \label{equ:T-matching-LP-vec}
\max \;\; \bm{Q}^T \bm{b} \quad \text{ s.t. } \bm{C}\bm{b} \le \bm{1}, \;\; \bm{b} \ge \bm{0}.
\ee

We now consider an example with $N=2$ and $T=3$. Constraint $\bm{C}\bm{b} \le \bm{1}$ can be shown as follows,
\be
\tiny{
\left(
  \begin{array}{c c c c c c c c c c c c}
    \red{1} & \red{1} & \red{0} & 0 & 0 & 0 & \red{0} & \red{0} & 0 & \red{0} & 0 & \red{0} \\
    0 & 0 & 1 & 1 & 0 & 0 & 0 & 0 & 0 & 0 & 0 & 0 \\
    \red{1} & \red{0} & \red{1} & 0 & 0 & 0 & \red{0} & \red{0} & 0 & \red{0} & 0 & \red{0} \\
    0 & 1 & 0 & 1 & 0 & 0 & 0 & 0 & 0 & 0 & 0 & 0 \\
    0 & 0 & 0 & 0 & 1 & 1 & 0 & 0 & 0 & 0 & 0 & 0 \\
    \red{0} & \red{0} & \red{0} & 0 & 0 & 0 & \red{1} & \red{1} & 0 & \red{0} & 0 & \red{0} \\
    0 & 0 & 0 & 0 & 1 & 0 & 1 & 0 & 0 & 0 & 0 & 0 \\
    0 & 0 & 0 & 0 & 0 & 1 & 0 & 1 & 0 & 0 & 0 & 0 \\
    0 & 0 & 0 & 0 & 0 & 0 & 0 & 0 & 1 & 1 & 0 & 0 \\
    0 & 0 & 0 & 0 & 0 & 0 & 0 & 0 & 0 & 0 & 1 & 1 \\
    0 & 0 & 0 & 0 & 0 & 0 & 0 & 0 & 1 & 0 & 1 & 0 \\
    \red{0} & \red{0} & \red{0} & 0 & 0 & 0 & \red{0} & \red{0} & 0 & \red{1} & 0 & \red{1} \\
    1 & 0 & 0 & 0 & 1 & 0 & 0 & 0 & 1 & 0 & 0 & 0 \\
    \red{0} & \red{1} & \red{0} & 0 & 0 & 1 & \red{0} & \red{0} & 0 & \red{1} & 0 & \red{0} \\
    \red{0} & \red{0} & \red{1} & 0 & 0 & 0 & \red{1} & \red{0} & 0 & \red{0} & 1 & \red{0} \\
    \red{0} & \red{0} & \red{0} & 1 & 0 & 0 & \red{0} & \red{1} & 0 & \red{0} & 0 & \red{1} \\
  \end{array}
\right)
\left(
  \begin{array}{c}
    b^1_{1,1} \\
    b^1_{1,2} \\
    b^1_{2,1} \\
    b^1_{2,2} \\
    b^2_{1,1} \\
    b^2_{1,2} \\
    b^2_{2,1} \\
    b^2_{2,2} \\
    b^3_{1,1} \\
    b^3_{1,2} \\
    b^3_{2,1} \\
    b^3_{2,2} \\
  \end{array}
\right)
\le
\left(
  \begin{array}{c}
    1 \\
    1 \\
    1 \\
    1 \\
    1 \\
    1 \\
    1 \\
    1 \\
    1 \\
    1 \\
    1 \\
    1 \\
    1 \\
    1 \\
    1 \\
    1 \\
  \end{array}
\right)
}
\nnb
\ee
Recall that a matrix is totally unimodular if the determinant of its any square submatrix is $0,1$ or $-1$.
Then let us take the following submatrix $\tilde{\bm{C}}$
with row indices $(1,3,6,12,14,15,16)$ and column indices $(1, 2, 3, 7, 8, 10, 12)$, i.e.,
\be
\tilde{\bm{C}}=
\left(
  \begin{array}{ccccccc}
    1 & 1 & 0 & 0 & 0 & 0 & 0 \\
    1 & 0 & 1 & 0 & 0 & 0 & 0 \\
    0 & 0 & 0 & 1 & 1 & 0 & 0 \\
    0 & 0 & 0 & 0 & 0 & 1 & 1 \\
    0 & 1 & 0 & 0 & 0 & 1 & 0 \\
    0 & 0 & 1 & 1 & 0 & 0 & 0 \\
    0 & 0 & 0 & 0 & 1 & 0 & 1 \\
  \end{array}
\right)
\ee
It turns out that the determinate of $\tilde{\bm{C}}$ is $-2$. Thus $\bm{C}$ is not totally unimodular.

This example shows that we cannot solve ILP \eqref{equ:T-matching-ILP} by solving its relaxed LP.
\fi

\bibliographystyle{IEEEtran}
\bibliography{ref}

 \begin{IEEEbiography}[{\includegraphics[width=1in,height=1.25in,clip,keepaspectratio]{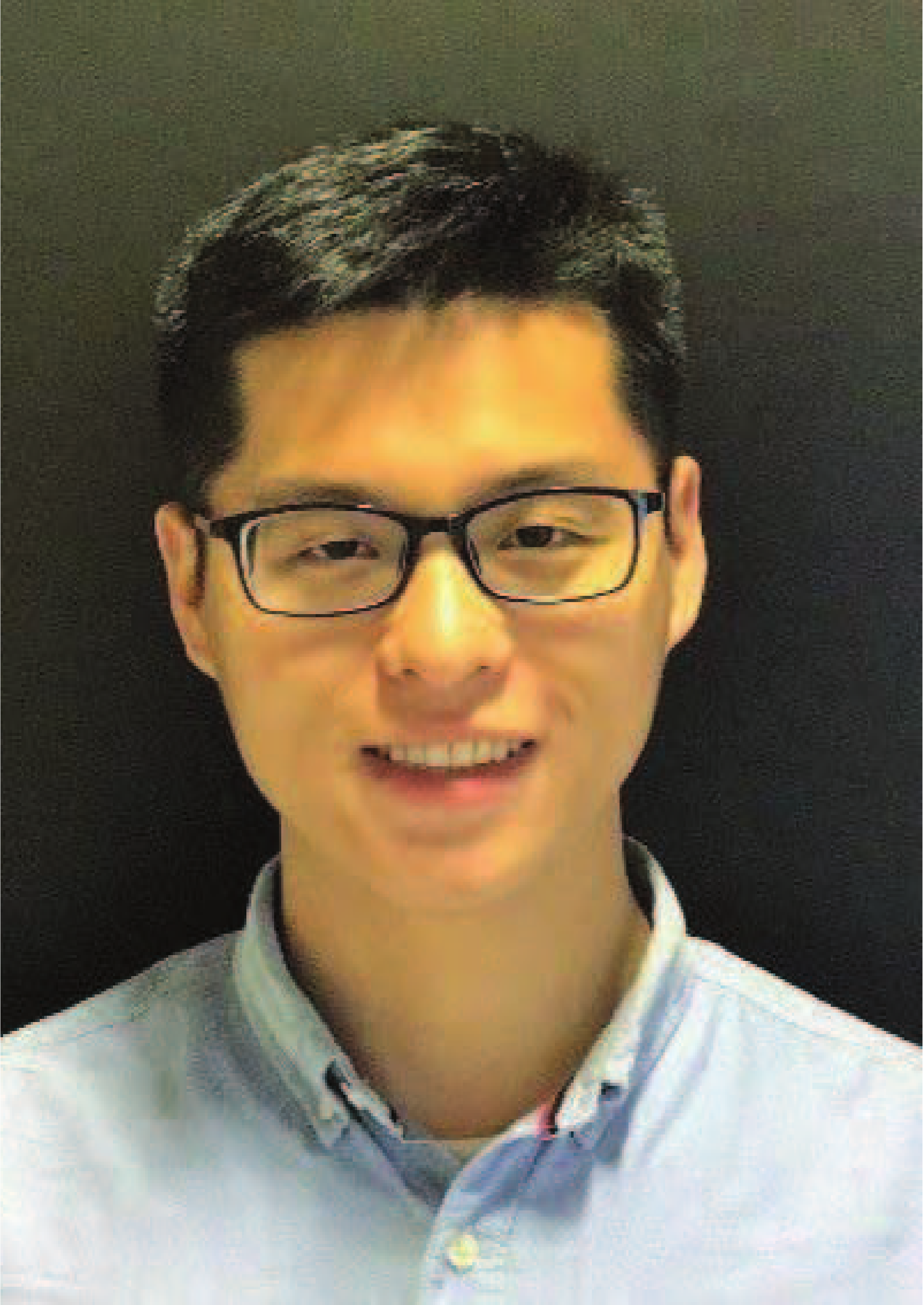}}]{Lei Deng}(M'17)
 received the B.Eng. degree from the Department of Electronic Engineering, Shanghai Jiao Tong University, Shanghai, China, in 2012, and the Ph.D. degree from the Department of Information Engineering, The Chinese University of Hong Kong, Hong Kong, in 2017. In 2015, he was a Visiting Scholar with the School of Electrical and Computer Engineering, Purdue University, West Lafayette, IN, USA. He is now an assistant professor in School of Electrical Engineering \& Intelligentization, Dongguan University of Technology. His research interests are timely network communications, intelligent transportation system, and spectral-energy efficiency in wireless networks.
\end{IEEEbiography}

\vspace{-1cm}

\begin{IEEEbiography}[{\includegraphics[width=1in,height=1.25in,clip,keepaspectratio]{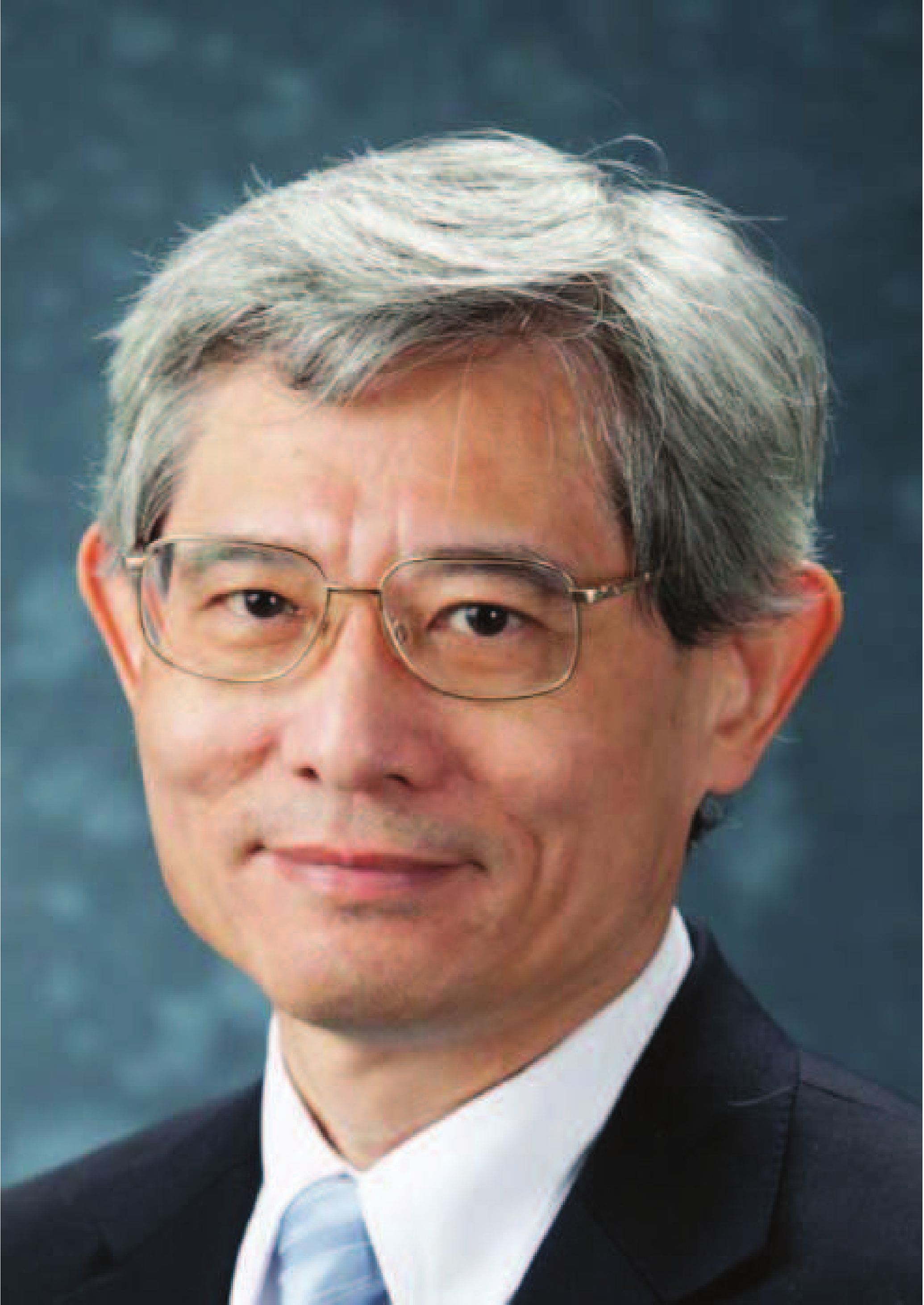}}]{Wing Shing Wong}(M'81--SM'90--F'02) received a combined master and bachelor degree from Yale University and M.S. and Ph.D. degrees from Harvard University.
He worked for the AT\&T Bell Laboratories from 1982 until he joined the Chinese University of Hong Kong in 1992, where he is now Choh-Ming Li Research Professor of Information Engineering. He was the Chairman of the Department of Information Engineering from 1995 to 2003 and the Dean of the Graduate School from 2005 to 2014.  He served as Science Advisor at the Innovation and Technology Commission of the HKSAR government from 2003 to 2005.  He has participated in a variety of research projects on topics ranging from mobile communication, networked control to network control.
\end{IEEEbiography}

\vspace{-1cm}

\begin{IEEEbiography}[{\includegraphics[width=1in,height=1.25in,clip,keepaspectratio]{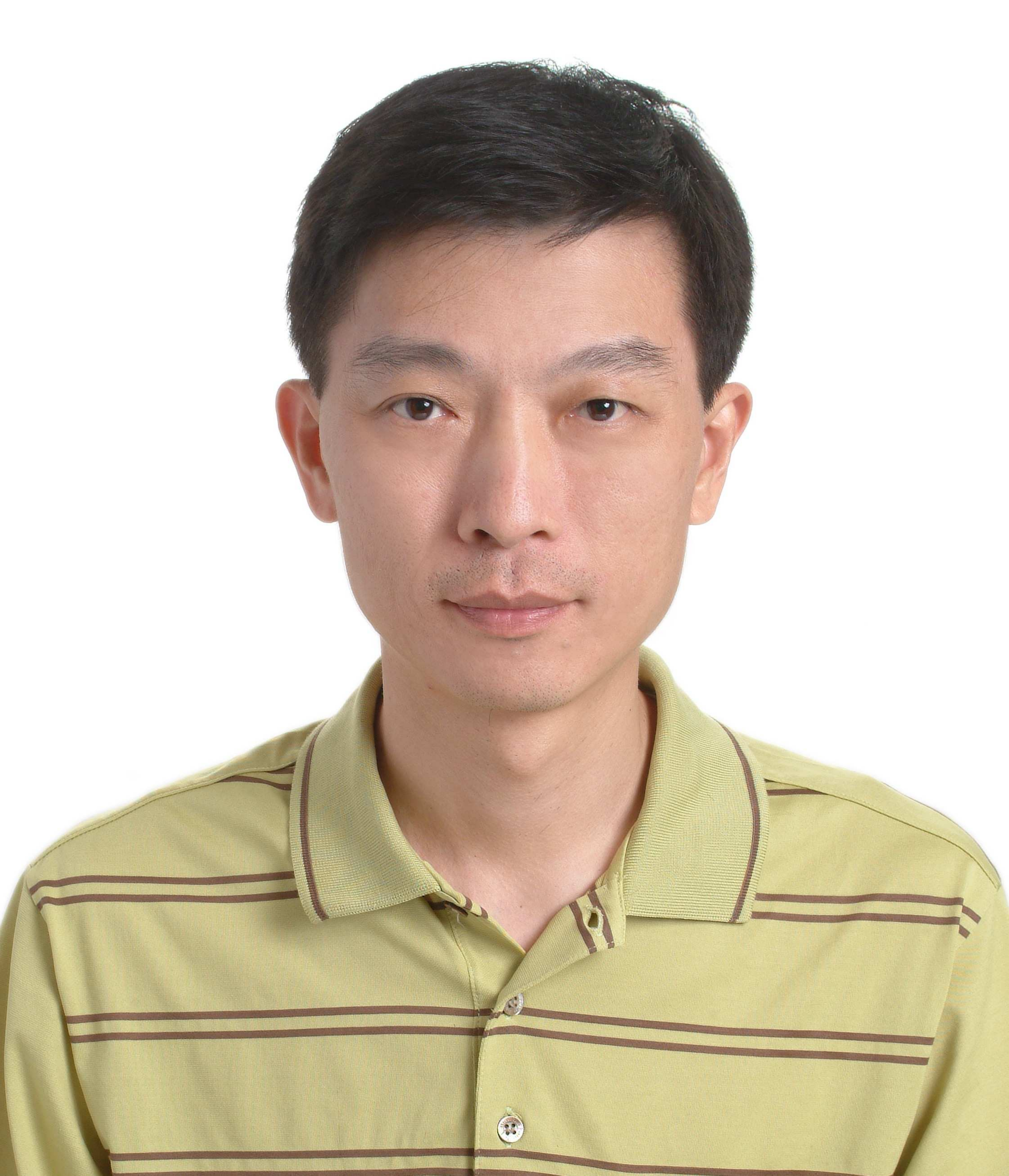}}]{Po-Ning Chen}(S'93--M'95--SM'01) received
  the Ph.D.\ degree in electrical engineering from University of
  Maryland, College Park, in 1994.
  Since 1996, he has been an Associate Professor in Department of
  Communications Engineering at National Chiao Tung University (NCTU),
  Taiwan, and was promoted to a full professor in 2001.
 He has served as the chairman
  of Department of Communications Engineering, NCTU, during
  2007--2009. From 2012--2015, he was the associate chief director of
  Microelectronics and Information Systems Research Center, NCTU, and is now the associate dean of the College of Electrical and Computer Engineering, NCTU.
  Dr.~Chen
  received the 2000 Young Scholar Paper Award from Academia Sinica,
  Taiwan. His research
  interests generally lie in information and coding theory, large
  deviations theory, distributed detection and sensor networks.
\end{IEEEbiography}

\vspace{-1cm}

\begin{IEEEbiography}[{\includegraphics[width=1in,height=1.25in,clip,keepaspectratio]{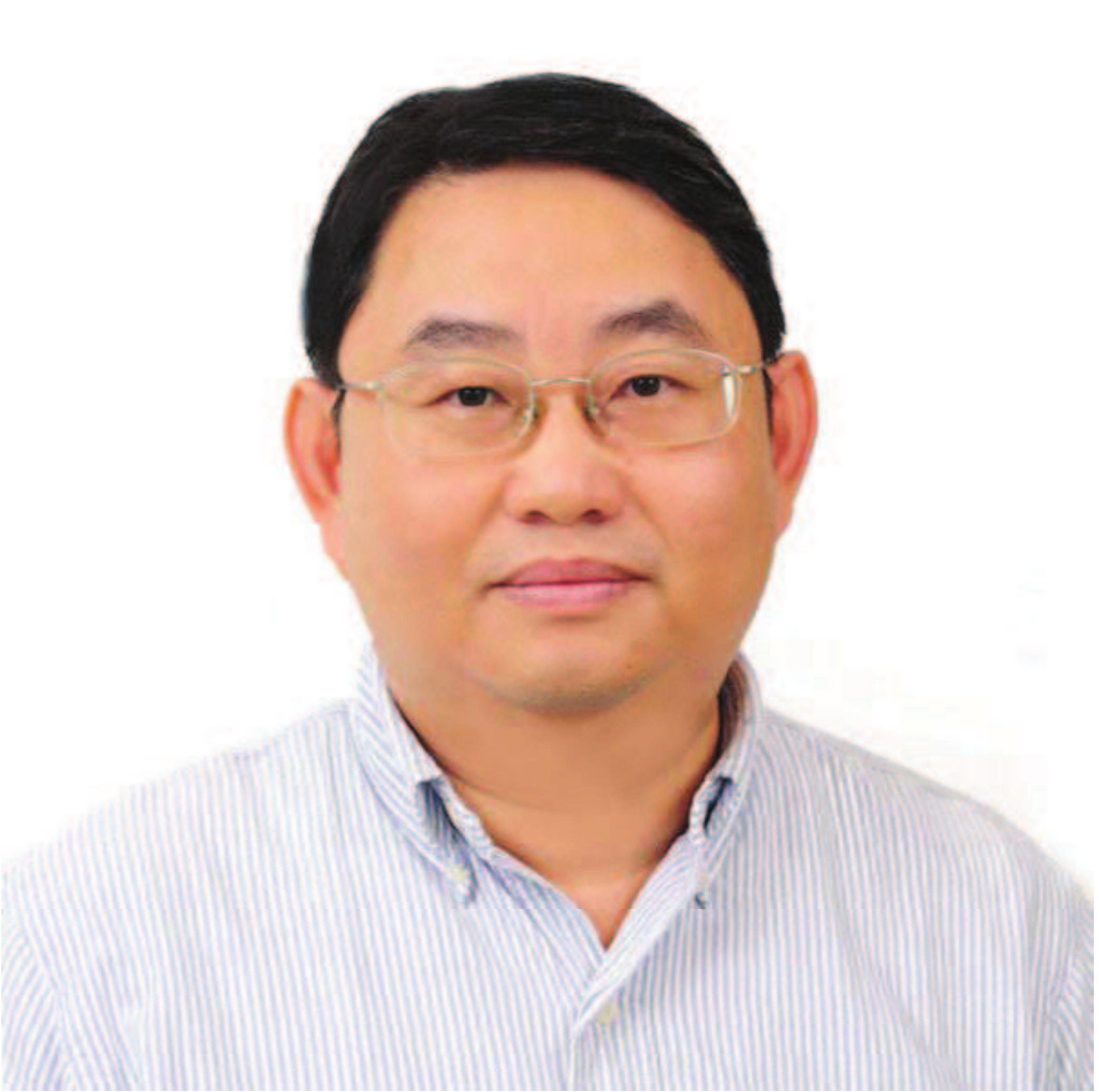}}]{Yunghsiang S. Han}(S'90-M'93-SM'08-F'11) was born in Taipei, Taiwan, 1962. He received B.Sc. and M.Sc. degrees in electrical engineering from the National Tsing Hua University, Hsinchu, Taiwan, in 1984 and 1986, respectively, and a Ph.D. degree from the School of Computer and Information Science, Syracuse University, Syracuse, NY, in 1993.
He is now with School of Electrical Engineering \& Intelligentization at Dongguan University of Technology, China. He is also a Chair Professor at National Taipei University from February 2015. His research interests are in error-control coding, wireless networks, and security.
Dr. Han was a winner of the 1994 Syracuse University Doctoral Prize and a Fellow of IEEE. One of his papers won the prestigious 2013 ACM CCS Test-of-Time Award in cybersecurity.
\end{IEEEbiography}

\vspace{-1cm}

\begin{IEEEbiography}[{\includegraphics[width=1in,height=1.25in,clip,keepaspectratio]{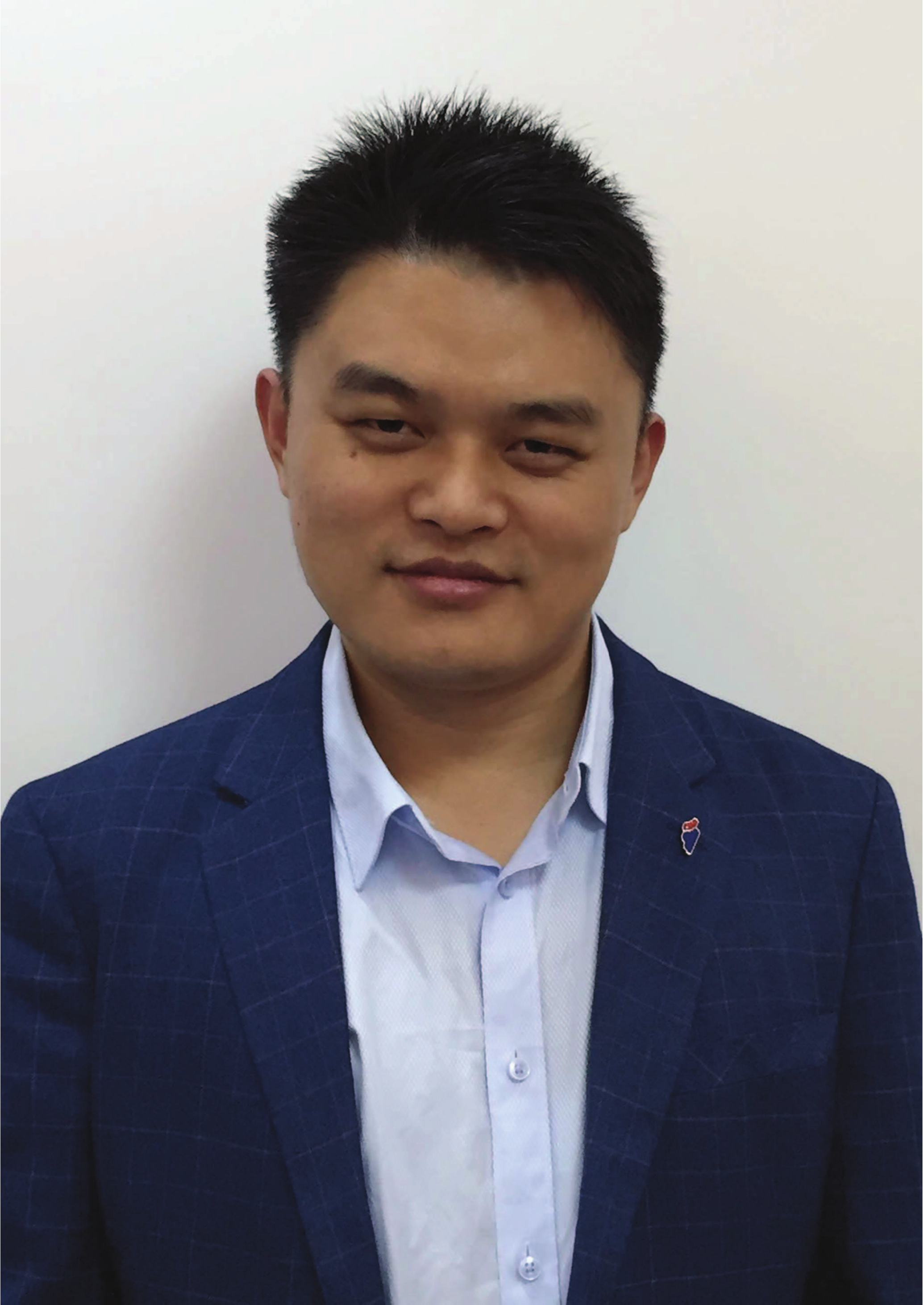}}]{Hanxu Hou}(S'11-M'16) was born in Anhui, China, 1987. He received the B.Eng. degree in Information Security from Xidian University, Xi’an, China, in 2010, and Ph.D. degrees in the Dept. of Information Engineering from The Chinese University of Hong Kong in 2015 and in the School of Electronic and Computer Engineering from Peking University in 2016. He is now an Assistant Professor with the School of Electrical Engineering \& Intelligentization, Dongguan University of Technology. His research interests include erasure coding and coding for distributed storage systems.
\end{IEEEbiography}

\end{document}